\documentclass[lettersize,journal,twoside]{IEEEtran}
\usepackage{amsmath}
\usepackage{amsfonts}
\usepackage{etoolbox}
\patchcmd{\subequations}{}%
{}{}{}

\usepackage[ruled, algonl, vlined]{algorithm2e}
\SetAlFnt{\footnotesize}
\SetAlCapFnt{\footnotesize}
\SetAlCapNameFnt{\footnotesize}

\usepackage{array}
\usepackage{booktabs}
\usepackage[caption=false,font=normalsize,labelfont=sf,textfont=sf]{subfig}
\usepackage{cite}
\usepackage{enumitem}
\usepackage{float}
\usepackage{graphicx}
\usepackage{multirow}
\usepackage{nomencl}
\usepackage{stfloats}
\usepackage{setspace}
\usepackage{textcomp}
\usepackage{tabularx}
\usepackage{url}
\usepackage{verbatim}
\usepackage{makecell}
\usepackage{dcolumn}

\usepackage{hyperref}  
\hypersetup{colorlinks = true,
            linkcolor  = red,
            citecolor  = red,
            urlcolor   = red}

\usepackage{colortbl}
\definecolor{LBlue}{RGB}{44, 77, 118}
\definecolor{CBlue}{RGB}{0, 0, 0}

\newtheorem{definition}{Definition}

\begin{document}
\allowdisplaybreaks

\title{\LARGE
A Carryover Storage {Valuation} Framework for {Medium-Term} Cascaded Hydropower Planning: A Portland General Electric System Study}
\author{Xianbang Chen,~\IEEEmembership{Student Member,~IEEE,}~Yikui Liu,~\IEEEmembership{Member,~IEEE,}~Zhiming Zhong,~Neng Fan,\\~Zhechong Zhao,~Lei Wu,~\IEEEmembership{Fellow,~IEEE}
\vspace{-10mm}
\thanks{
This work is supported in part by the U.S. Department of Energy’s Office of Energy Efficiency
and Renewable Energy (EERE) under the Water Power Technologies Office Award Number DE-EE0008944.
The views expressed herein do not necessarily represent the views of the U.S. Department of Energy or the
United States Government.

X. Chen and L. Wu are with the ECE Department, Stevens Institute of Technology, Hoboken, NJ, 07030 USA. (Email: xchen130@stevens.edu; lei.wu@stevens.edu)

Y. Liu is with the Electrical Engineering Department, Sichuan University, Chengdu, 610017 China. (Email: yikuiliu@gmail.com)

Z. Zhao is with Portland General Electric, Portland, OR, 97204 USA. (Email: zhechong.zhao@pgn.com)

Z. Zhong and N. Fan are with the Department of Systems and Industrial Engineering, University of Arizona, Tucson, AZ, 85721 USA. (Email: zhongz@arizona.edu; nfan@arizona.edu)
}}

\maketitle

\begin{abstract}
Medium-term planning of cascaded hydropower (CHP) determines appropriate carryover storage levels in reservoirs to optimize the usage of available water resources. This optimization seeks to maximize the hydropower generated in the current period (i.e., immediate benefit) plus the potential hydropower generation in the future period (i.e., future value). Thus, in the medium-term CHP planning, properly quantifying the future value deposited in carryover storage is essential to achieve a balanced trade-off between immediate benefit and future value. To this end, this paper presents a framework to quantify the future value of carryover storage, which consists of three major steps: \textit{i)} constructing a model to calculate the maximum possible hydropower generation that a given level of carryover storage can deliver in the future period; \textit{ii)} extracting the implicit locational marginal water value (LMWV) of carryover storage for each reservoir by applying a partition-then-extract algorithm to the constructed model; and \textit{iii)} developing a set of analytical rules based on the extracted LMWV to effectively calculate the future value. These rules can be seamlessly integrated into medium-term CHP planning models as tractable mixed-integer linear constraints to quantify the future value properly, and can be easily visualized to offer valuable insights for CHP operators. Finally, numerical results on a CHP system of Portland General Electric demonstrate the effectiveness of the presented framework in determining proper carryover storage values to facilitate medium-term CHP planning.

\end{abstract}
\vspace{-0mm}

\vspace{-0mm}
\section*{Nomenclature}
\vspace{-0mm}
\addcontentsline{toc}{section}{Nomenclature}
\begin{spacing}{1.05}
{\color{CBlue}Major symbols used throughout the paper are defined here, while others are clarified as needed after their first appearance.}
\subsection*{Sets and Indices:}
\begin{IEEEdescription}[\IEEEusemathlabelsep \IEEEsetlabelwidth{$\mspace{45mu}$} \setlength{\IEEElabelindent}{0pt}]

\item[$\mathcal{\tilde{B}}_{h}$]
Set of explored binaries for critical region (CR) $h$.

\item[$\mathcal{I}_{n}$]
Set of units in reservoir $n$, indexed by $i$.

\item[$\mathcal{L}/l$]
Set/index of weeks covered in the future period, i.e., $l \in \mathcal{L}=\{T+1,..., T+L\}$. 

\item[$\mathcal{N}/n$]
Set/index of reservoirs, i.e., $n \in \mathcal{N} = \{1,..., N\}$.

\item[$\bar{\mathcal{N}}_{n}/m$]
Set/index of direct upstream reservoirs of reservoir $n$.

\item[$\mathcal{R}/r, h$]
Set/indices of CRs, i.e., $r, h \in \mathcal{R}=\{1,..., R\}$.

\item[$\mathcal{T}/t,\tau$]
Set/indices of weeks covered in the current period, i.e., $t, \tau \in \mathcal{T}=\{1,..., T\}$.

\end{IEEEdescription}

\vspace{-0mm}
\subsection*{Decision Variables:}
\vspace{-0mm}
\begin{IEEEdescription}[\IEEEusemathlabelsep \IEEEsetlabelwidth{$\mspace{45mu}$} \setlength{\IEEElabelindent}{0pt}]
\item[$D_{ni}$]
Discharge rate of unit $i$ on reservoir $n$. [{\color{CBlue}m\textsuperscript{3}/s}]

\item[$I_{ni}$]
ON-OFF status of unit $i$ on reservoir $n$.

\item[$L_{n}^{\text{n-dis}/\text{dis}}$]
Length of non-discharge/discharge phase of reservoir $n$ in the future period. [week]

\item[$L_{vu}^{\Delta}$]
Gap between discharge phases of reservoirs $v$ and $u$. [week]

\item[$P_{ni}$]
Hydropower dispatch of unit $i$ on reservoir $n$. [MW]

\item[$S_{n}$]
Water spillage of reservoir $n$. [{\color{CBlue}Mm\textsuperscript{3}}]

\item[$V_{n}^{\text{cs}}$]
Carryover storage of reservoir $n$, forming vector $\boldsymbol{V}^{\text{cs}}$. [{\color{CBlue}Mm\textsuperscript{3}}]

\item[$W_{n}^{\Delta}$]
Water difference between the total inflows and outflows of reservoir $n$. [{\color{CBlue}Mm\textsuperscript{3}}]

{\color{CBlue}
\item[$W_{vn}^{\text{i/o}}$]
Water inflow/outflow from direct upstream reservoir $v$ to reservoir $n$ during $L_{vu}^{\Delta}$.
}

\item[$Z_{r}$]
Binary indicator for CR $r$.

{\color{CBlue}
\item[$\diamond$]
Superscript to indicate variables of the current period.

}

\end{IEEEdescription}

\vspace{-4mm}
\subsection*{Parameters:}
\vspace{-0mm}
\begin{IEEEdescription}[\IEEEusemathlabelsep \IEEEsetlabelwidth{$\mspace{45mu}$} \setlength{\IEEElabelindent}{0pt}]
\item[$C_{n}^{\text{ws}}$]
Water spillage penalty of reservoir $n$. [MWh/{\color{CBlue}Mm\textsuperscript{3}}]

\item[$D_{ni}^{\text{M}/\text{m}}$]
Max/min discharge rate of unit $i$ on reservoir $n$. [{\color{CBlue}m\textsuperscript{3}/s}]

\item[$\boldsymbol{e}_h, f_{h}$ ]
Constant vector and scalar to describe CR $h$.

\item[$K$]
Number of samples for the Bayesian neural network.

\item[$L/T$]
Length of future/current period. [week]

\item[$O_{r/h}$] Number of inequalities to express CR $r/h$.

\item[$P_{ni}^{\text{M}/\text{m}}$]
Max/min power limit of unit $i$ on reservoir $n$. [MW]

\item[$R$]
Number of final CRs.

\item[$V_{n}^{\text{cs},\theta}$]
The counterpart of $V_{n}^{\text{cs}}$ representing parameters in the multi-parametric programming model.

\item[$V_{n}^{\text{M}/\text{m}}$]
Max/min storage level of reservoir $n$. [{\color{CBlue}Mm\textsuperscript{3}}]

\item[$\hat{\boldsymbol{W}}^\text{cp/fp}$]
Water inflow prediction for the current/future period.

\item[$\pi_{rn}$]
LMWV of reservoir $n$ in CR $r$. [MWh/{\color{CBlue}Mm\textsuperscript{3}]}

{\color{CBlue}\item[$\lambda,\alpha$]
Unit-converting parameters, set as 168 h/week and 0.6048 Mm\textsuperscript{3}$\cdot$s/m\textsuperscript{3}, respectively.}

\end{IEEEdescription}

\vspace{-4mm}
\subsection*{Functions and Regions:}
\vspace{-0mm}
\begin{IEEEdescription}[\IEEEusemathlabelsep \IEEEsetlabelwidth{$\mspace{45mu}$} \setlength{\IEEElabelindent}{0pt}]

{\color{CBlue}\item[$A(\cdot)$]
Function for calculating immediate benefit. [MWh]}

\item[$F(\cdot)$]
Function for converting carryover storage (in {\color{CBlue}m\textsuperscript{3}/s}) into future value (in MWh).

\item[$\mathcal{P}^{\text{RtP}}(\cdot)$]
Piece-wise linearized function converting discharge rate (in {\color{CBlue}m\textsuperscript{3}/s}) into power (in MW).

\item[$\mathcal{V}^{\text{E}}$] 
Entire feasible region of carryover storage.

\item[$\mathcal{V}_{r}^{\text{CR}}$]
CR $r$ that is one of the final $R$ CRs. $\mathcal{V}^{\text{CR}}_{r} \subseteq \mathcal{V}^{\text{E}}$.

\item[$\hat{\mathcal{V}}^{\text{CR}},\mathcal{V}^{\text{P}}$]
CRs that could potentially be partitioned further.

\end{IEEEdescription}

{\color{CBlue}
\vspace{-4mm}
\subsection*{Abbreviations:}
\vspace{-0mm}
\begin{IEEEdescription}[\IEEEusemathlabelsep \IEEEsetlabelwidth{$\mspace{45mu}$} \setlength{\IEEElabelindent}{0pt}]
\item[BMDN]
Bayesian mixture density network

\item[CCP]
Chance-constrained programming

\item[CHP]
Cascaded hydropower

\item[CI]
Confidence interval

\item[CR]
Critical region

\item[DNN]
Deep neural network

\item[GMM]
Gaussian mixture model

\item[ICC]
Individual chance constraint

\item[IPLB]
Incumbent parametric lower bound

\item[JCC]
Joint chance constraint

\item[LMWV]
Locational marginal water value

\item[mp-LP]
Multi-parametric linear programming

\item[MAE]
Mean absolute error

\item[MAPE]
Mean absolute percentage error

\item[MILP]
Mixed-integer linear programming

\item[NSE]
Nash–Sutcliffe model efficiency

\item[PGE]
Portland General Electric

\item[PT]
Pelton

\item[RB]
Round butte

\item[RMSE]
Root mean square error

\item[WI]
Water inflow
\end{IEEEdescription}
}
\end{spacing}

\vspace{-0mm}
\section{Introduction}
\vspace{-0mm}
\subsection{Background}
\IEEEPARstart{I}{n} the U.S. electricity grid, hydropower fleets are valued as one of the golden assets because of their rapid responsiveness and environmental benefits. Among these hydropower technologies, cascaded hydropower (CHP) stands out for its remarkable efficiency \cite{CHSIntro}.

\begin{figure}[b]
	\centering
 \vspace{-4mm}
		\includegraphics[width=\columnwidth]{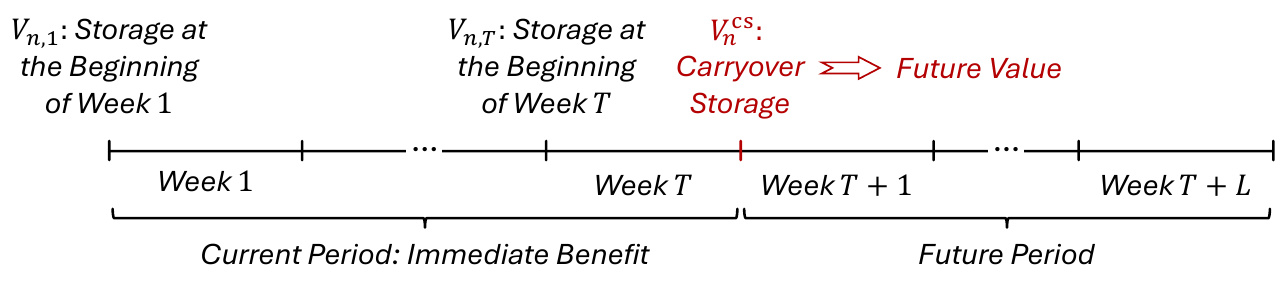}
  \vspace{-5mm}
	\caption{Illustration of medium-term CHP planning.}\label{Fig01}
\end{figure}

To manage CHP operations, CHP operators need to coordinate hierarchical decision-making tasks across multiple timescales, including the short-term scheduling problem to determine hydropower generation plans and the medium-term planning problem to provide water usage guidance for short-term operations. {\color{CBlue} The authors' previous work \cite{Predict_Yikui} presents a decision-making framework for short-term CHP scheduling. In comparison, this paper focuses on the medium-term CHP planning problem to determine proper carryover storage values for guiding short-term scheduling \cite{STandMT}.} As shown in Fig.~\ref{Fig01}, the medium-term planning horizon covers a \textit{current period} that focuses on the scheduling of upcoming $T$ weeks, and a subsequent \textit{future period} that looks ahead to $L$ weeks immediately following the current period. In this paper, the hydropower generation during the current period is referred to as \textit{immediate benefit}, while the hydropower generation that the carryover storage can deliver in the future period is referred to as the \textit{future value}. The medium-term planning aims to determine an optimal carryover storage level $V^{\text{cs}}_{n}$, i.e., the remaining water volume in reservoir $n$ at the end of the current period, that can maximize the total immediate benefit and future value by balancing how much water to use in the current period versus how much to hold for future use.

Indeed, one critical issue is how to properly quantify the future value in the medium-term CHP planning problems.
Specifically, overestimating the future value (i.e., overestimating the hydropower generation (in MWh) that per unit of stored water can deliver in the future period) would lead to excessive carryover storage, potentially causing unnecessary water spillage in the future; conversely, underestimating the future value may result in insufficient carryover storage, potentially causing hydropower supply shortages in the future.

For instance, in 1983, the operators of Glen Canyon Dam in Arizona overestimated the future value, leading to excessively high carryover storage. With this high carryover storage and an unexpectedly large water inflow (WI) that occurred in the spring of 1984, the operators were forced to spill excessive water to prevent overtopping. This action caused severe spillway erosion and significant water waste \cite{Glen}. As another example, in 2021, an underestimation of the future value led to excessive water usage from Shasta Dam in California during the current period. This mismanagement resulted in critically low reservoir storage later in the year, severely affecting the dam's power supply to Northern California \cite{Shasta}.

More recently, the National Integrated Drought Information System reported that, as of mid-2024, over 24\% of Oregon experienced precarious WI conditions \cite{Oregon}. Such conditions pose significant challenges to hydropower systems, including the CHP operated by Portland General Electric (PGE), underscoring the need for a more delicate quantification of future value to effectively manage water usage over the medium-term horizon. To this end, this paper is dedicated to the following question {\color{CBlue} from the perspective of CHP operators}: \textit{How can the future value of CHP carryover storage be properly quantified in an interpretable, hydrological adaptive, and easy-to-use manner?} By properly determining how much hydropower the carryover storage can deliver in the future period, the medium-term CHP planning problem can yield more informed carryover storage decisions, as demonstrated through numerical simulations on PGE's CHP system.

\vspace{-0mm}
\subsection{Literature Review}
\vspace{-0mm}
The core of quantifying future value is a concept called \textit{locational marginal water value (LMWV)} \cite{SDP_Helseth}, representing hydropower (in MWh) that a reservoir can produce with one incremental unit of stored water. The term ``locational'' emphasizes the different effects of individual reservoirs.
To ensure the proper deployment of LMWV in the medium-term planning problem, significant efforts have been made, which can be broadly categorized into three types of methods:

\begin{itemize}
{\color{CBlue}\item[\textit{i)}] The first type of method uses rule-of-thumb approaches to calculate LMWVs \cite{type1_a, type1_b, type1_c}. For example, reference \cite{type1_a} employs a heuristic rule to calculate LMWVs as a weighted sum of recent electricity prices via stochastically generated utility coefficients.} With this, the future value can be calculated as a linear function, i.e., the product of constant LMWV and carryover storage variable, in the medium-term planning problem. Although intuitive and easy-to-use, the constant LMWVs suffer from three issues. First, they are heavily experience-dependent, which may differ significantly across experts. Second, LMWVs should vary against dynamic hydrological factors but are ignored in the constant method; for instance, near-full carryover storage reduces the urgency to conserve water and shall be associated with small LMWVs, while lower future WIs make storage more critical for future use and shall be accompanied with large LMWVs. Last, these LMWVs are generally inferred from historical patterns, making them less adaptable to rare or precarious WI conditions;

\item[\textit{ii)}] The second type of method, e.g., \cite{type2_b, type2_c, type2_d}, trains data-driven models to predict LMWVs. This method can learn the complex and dynamic relationship between LMWVs and hydrological factors. However, the training process often involves heuristic steps, leading to randomness and inconsistencies in the trained models. For instance, even with the same set of data and hyperparameters, the trained models may still have different parameters across individual training runs. More importantly, it is difficult for CHP operators to directly interpret LMWV results generated by the data-driven method;

{\color{CBlue}
\item[\textit{iii)}]
The third type of method relies on rigorous mathematical methodologies (e.g., dual theory). Within this category, stochastic dynamic programming (SDP) has been shown to be well-suited for calculating LMWVs \cite{SDP_Helseth}. However, SDP becomes computationally taxing when applied to multiple reservoirs and/or longer time periods because it requires discretizing state variables (e.g., reservoir storage levels). To mitigate computational challenges, stochastic dual dynamic programming (SDDP) \cite{SDDP} has been applied in the CHP field \cite{Arild1, Arild_aggregation, Arild6, Arild7, Arild_TPS, binaryexpansion}. The SDDP-based methods are generally built on the definition that, at a certain level of carryover storage, dual variables of the water balance constraints represent the corresponding LMWVs. With this, Benders cuts are constructed to provide a convex approximation of the future value function. A notable advantage of these cuts is their physical interpretability: each one is a linear inequality formed by LMWVs multiplied by the carryover storage variables. However, SDP-based methods rely on discretized approximations, and SDDP-based methods provide only convex approximations, which cannot precisely capture the future value surface, especially when inherent non-convexity (e.g., discontinuities caused by binary variables indicating hydro unit ON-OFF statuses) is involved.
}
\end{itemize}

\vspace{-0mm}
\subsection{The Proposed Work and Major Contributions}
\vspace{-0mm}
Within the field of medium-term CHP planning, this paper introduces a quantification framework to address the following question {\color{CBlue}from the perspective of CHP operators}: \textit{How can the future value of CHP carryover storage be properly quantified in an interpretable, hydrological adaptive, and easy-to-use manner?} The proposed framework comprises the following three major steps:

\begin{itemize}
\item[\textit{i)}]
Building a model to calculate the maximum possible hydropower generation over the future period using two key inputs: \textit{a)} data-driven WI predictions for the future period, and \textit{b)} carryover storage at the end of the current period, which is treated as an unspecified parameter;

\item[\textit{ii)}]
Applying a partition-then-extract algorithm to the future period model from step \textit{i)}, in order to extract a set of unique LMWV vectors corresponding to different ranges of carryover storage;

\item[\textit{iii)}]
Based on the LMWVs extracted in step \textit{ii)}, a set of analytical rules is formulated to calculate the future value. These rules allow CHP operators to quantify future value in a straightforward (using simple multiplication), interpretable (grounded in well-established dual theory), and hydrologically adaptive (accounting for carryover storage and future WI) way. Moreover, these rules can be seamlessly integrated into a medium-term CHP planning model as tractable linear constraints.
\end{itemize}

{\color{CBlue}
In summary, this paper makes the following contributions:
\begin{itemize}
\item[\textit{i)}]
A quantification framework, comprising the future period model and the partition-then-extract algorithm, is developed to provide CHP operators with an analytical alternative for analyzing future value. Compared to rule-of-thumb (e.g., \cite{type1_a, type1_b, type1_c}) and data-driven (e.g., \cite{type2_b, type2_c, type2_d}) methods, the proposed framework can offer quantification results with stronger mathematical interpretability;

\item[\textit{ii)}]
Compared to SDP-based methods (e.g., \cite{SDP_Helseth}) and SDDP-based methods (e.g., \cite{Arild1, Arild_aggregation, Arild6, Arild7, Arild_TPS, binaryexpansion}), the presented partition-then-extract algorithm offers a distinct advantage that the derived analytical rules can precisely capture the entire future value surface. This precise representation provides CHP operators with comprehensive geometric insights into the future value;

\item[\textit{iii)}]
Numerical simulations based on the PGE's CHP system and a modified larger 8-reservoir CHP are thoroughly conducted and analyzed. The visualization results demonstrate that the presented framework can assist CHP operators, such as PGE, in improving their current practice for quantifying future value and improving medium-term CHP planning.
\end{itemize}
}

The rest of the paper is organized as follows: Section~\ref{Sec2} introduces preliminaries; Section~\ref{Sec3} elaborates on the presented framework; Section~\ref{Sec4} analyzes the numerical cases on PGE's CHE; and Section~\ref{Sec5} concludes this paper. 

\vspace{-0mm}
\section{Preliminaries}\label{Sec2}
\vspace{-0mm}
\subsection{General Formulation of Medium-Term CHP Planning}
\vspace{-0mm}
In general, CHP operators can determine the carryover storage by solving a medium-term CHP planning problem as in \eqref{General}. The objective function \eqref{General:1} is to maximize the summation of the immediate benefit and the future value.

The current period decisions include $\boldsymbol{V}^{\text{cs}}$ (the carryover storage) and $\Xi$ (remaining current period decisions, e.g., discharging). Their feasible region $\mathcal{X(\hat{\boldsymbol{W}}^\text{cp})}$, bounded by the WI prediction $\hat{\boldsymbol{W}}^\text{cp}$ of the current period as \eqref{General:2}, includes prevalent hydropower scheduling constraints such as storage limits, discharge limits, and water balance requirements.

The future value function $F(\cdot)$ takes the carryover storage $\boldsymbol{V}^{\text{cs}}$ and the WI prediction $\hat{\boldsymbol{W}}^\text{fp}$ of the future period as inputs.\begin{subequations}\label{General}
\begin{align}
\max_{\boldsymbol{V}^{\text{cs}}, \Xi}&
\overbrace{A(\boldsymbol{V}^{\text{cs}}, \Xi, \hat{\boldsymbol{W}}^{\text{cp}}         )}^{\text{Immediate Benefit}} +
\overbrace{F(\boldsymbol{V}^{\text{cs}}, \hat{\boldsymbol{W}}^{\text{fp}})}^{\text{Future Value}}    \label{General:1}\\
\text{s.t. }  &\boldsymbol{V}^{\text{cs}}, \Xi  \in \mathcal{X}(\hat{\boldsymbol{W}}^{\text{cp}});   \label{General:2}
\end{align}
\end{subequations}

\vspace{-0mm}
{\color{CBlue}
\subsection{Benefits of Using the Presented Quantification Framework}
Before elaborating on the presented quantification framework, it is worth clarifying why CHP operators do not directly solve the medium-term planning model \eqref{General} to determine the carryover storage. Indeed, as directly solving \eqref{General} requires a non-trivial scheduling model to explicitly formulate $F(\cdot)$, the proposed quantification framework presents favorable features to CHP operators for two major reasons:
\begin{itemize}
\item
Typical CHP scheduling models calculate hydropower generation from both carryover storage and future WI. However, the future value $F(\cdot)$ specifically represents the portion of hydropower generation contributed solely by carryover storage. To exactly quantify this exclusive contribution from the total generation, specific methodologies, such as the presented framework, are required;

\item
The process of using optimization models to calculate the future value $F(\cdot)$ generally appears as a black box to CHP operators. Regarding this, a notable advantage of the presented framework is its clear interpretability rendered by analytical quantification rules.
\end{itemize}
}

\section{The Future Value Quantification Framework}\label{Sec3}
\vspace{-0mm}
The core of the presented framework is to derive LMWV vectors, enabling the future value function $F({\cdot})$ to be expressed in a linear form. The process includes three steps:
\begin{itemize}
\item
Section~\ref{Step1}: Build a model to calculate the maximum possible hydropower generation over the future period with respect to the carryover storage and expected future WI;

\item
Section~\ref{Step2}: Derive LMWV vectors from the model using the partition-then-extract algorithm;

\item
Section~\ref{Step3}: Use the derived LMWVs to construct a set of analytical ``if-then'' rules to express the future value function.
\end{itemize}

\subsection{Calculate Maximum Possible Generation in Future Period}\label{Step1}

\subsubsection{Build Model \eqref{FModel} to Calculate the Maximum Hydropower Generation}

{\color{CBlue}Model \eqref{FModel} is constructed to calculate the maximum possible hydropower generation over the future period. Specifically, model \eqref{FModel} approximates future CHP operations by aggregating potential operation actions of each reservoir $n$ into two sequential phases: a non-discharge phase lasting $L_{n}^{\text{n-dis}}$ weeks in which reservoir $n$ receives WIs but does not release water, 
and a subsequent discharge phase lasting the remaining $L_{n}^{\text{dis}}$ weeks (i.e., $L_{n}^{\text{n-dis}} +  L_{n}^{\text{dis}} = L$) in which it discharges water to generate hydropower while continuing receiving WIs.
Model \eqref{FModel} features that \textit{operators can jointly optimize how long and how much water each reservoir $n$ should discharge to achieve the maximum hydropower generation}, with respect to the uniform discharge rate variable $D_{ni}$.}
\begin{subequations}\label{FModel}
\begin{flalign}
& \textstyle{\max\limits_{\Psi}
\sum\nolimits_{n \in \mathcal{N}} \sum\nolimits_{i \in \mathcal{I}_{n}} L^{\text{dis}}_{n} \lambda P_{ni} -  \sum\nolimits_{n \in \mathcal{N}} C^{\text{ws}}_{n} S_{n}}                                    \mspace{-200mu} &                 \notag\\
&\text{where } \Psi =\{
\boldsymbol{D},
\boldsymbol{I},
\boldsymbol{L}^{\text{n-dis}/\text{dis}},
\boldsymbol{L}^{\Delta},
\boldsymbol{P},
\boldsymbol{S},
\boldsymbol{W}^{\Delta\text{/i/o}}\}                          \mspace{-200mu}&                        \label{FModel:1}\\
&\text{s.t. } \textstyle{L^{\text{n-dis}}_{n} + L^{\text{dis}}_{n} = L,\, L^{\text{n-dis}}_{n} \geq 0, \, L^{\text{dis}}_{n} \geq 0,}                                                   \mspace{-200mu}&\forall n;              \label{FModel:2}\\
&\mspace{2mu}
\renewcommand{\arraystretch}{1.5}
\left\{\mspace{-11mu} 
\begin{array}{lr}
V^{\text{m}}_{n} \leq  V_{n}^{\text{cs},\theta} + \frac{{L^{\text{n-dis}}_{v}}\hat{W}^\text{fp}_{n}}{L} + W^{\text{i}}_{vn} - W^{\text{o}}_{vn},                            &\mspace{-34mu}\forall v  \in \{n, \bar{\mathcal{N}}_{n} \};\\
V^{\text{M}}_{n} \geq V_{n}^{\text{cs},\theta} + \frac{{L^{\text{n-dis}}_{v}}\hat{W}^\text{fp}_{n}}{L} + W^{\text{i}}_{vn} - W^{\text{o}}_{vn},                            &\mspace{-34mu}\forall v  \in  \{n, \bar{\mathcal{N}}_{n} \};\\
W^{\text{i}}_{vn} = \sum\nolimits_{m \in \bar{\mathcal{N}}_{n} } ( L^{\Delta}_{vm} \sum\nolimits_{i \in \mathcal{I}_{m}} \alpha D_{mi}),                                  &\mspace{-34mu}\forall v  \in \{n, \bar{\mathcal{N}}_{n} \}; \\
W^{\text{o}}_{vn} = L^{\Delta}_{vn} \sum\nolimits_{i \in \mathcal{I}_{n}}\alpha D_{ni}, 
                                                           &\mspace{-34mu}\forall v \in \{n, \bar{\mathcal{N}}_{n} \};\\
                                                           L^{\Delta}_{vu} = \max\{0, L_{v}^{\text{n-dis}} - L_{u}^{\text{n-dis}}\},
                                                           &\mspace{-34mu}\forall v, u \in  \{n, \bar{\mathcal{N}}_{n}\};
\end{array}
\mspace{-11mu}\right\},                                                       \mspace{-200mu}&  \notag                  \\
&\mspace{9mu}                                                                \mspace{-200mu}&\forall n;              \label{FModel:3}\\
&\mspace{9mu} \textstyle{V^{\text{m}}_{n} \leq  V_{n}^{\text{cs},\theta} + \hat{W}^\text{fp}_{n} + W^{\Delta}_{n}
+ \sum\nolimits_{m \in \bar{\mathcal{N}}_{n} }S_{m} - S_{n},}                \mspace{-200mu}&\forall n;              \label{FModel:value}\\
&\mspace{9mu} \textstyle{V^{\text{M}}_{n} \geq V_{n}^{\text{cs},\theta} + \hat{W}^\text{fp}_{n} + W^{\Delta}_{n}
+ \sum\nolimits_{m \in \bar{\mathcal{N}}_{n} }S_{m} - S_{n},}                \mspace{-200mu}&\forall n;              \label{FModel:5}\\
&\mspace{9mu} \textstyle{W^{\Delta}_{n} = \sum\nolimits_{m \in \bar{\mathcal{N}}_{n}} ( L^{\text{dis}}_{m}\sum\nolimits_{i \in \mathcal{I}_{m}} \alpha D_{mi}) }    \mspace{-200mu}&                  \notag\\
&\mspace{9mu} \textstyle{\quad -  L^{\text{dis}}_{n}\sum\nolimits_{i \in \mathcal{I}_{n}} \alpha D_{ni},\,S_{n} \geq 0,}                                      \mspace{-200mu}&\forall n;              \label{FModel:6}\\
&\mspace{9mu} \textstyle{P_{ni}=\mathcal{P}^{\text{RtP}}(D_{ni}, I_{ni}), I_{ni} \in \{0, 1\},}
                                                                             \mspace{-200mu}&\forall n, \forall i;   \label{FModel:7}\\
&\mspace{9mu} \textstyle{P^{\text{m}}_{ni} I_{ni} \leq P_{ni} \leq  P^{\text{M}}_{ni} I_{ni},}
                                                                             \mspace{-200mu}&\forall n, \forall i;   \label{FModel:8}\\
&\mspace{9mu} \textstyle{D^{\text{m}}_{ni} I_{ni} \leq D_{ni}  \leq D^{\text{M}}_{ni} I_{ni},}
                                                                             \mspace{-200mu}&\forall n, \forall i;   \label{FModel:9}
\end{flalign}
\end{subequations}

The objective function \eqref{FModel:1} maximizes the hydropower generation (in MWh) minus water spillage penalization during the future period. {\color{CBlue} The water spillage variables \(S_n\) act as slack variables to ensure the feasibility of model \eqref{FModel}.}

Constraint \eqref{FModel:2} indicates that each reservoir involves one non-discharge phase and one discharge phase. As individual reservoirs would start discharging asynchronously, a group of constraints, as described in \eqref{FModel:3}, is designed to describe the storage limit of each reservoir $n$ over the $L$ weeks. The logic is that a phase switch (from non-discharge to discharge) for any reservoir triggers a storage limit check.

In \eqref{FModel:3}, the first and second inequalities enforce the storage limits; the third equation calculates the WI from the direct upstream reservoirs to reservoir $n$ during $L^{\Delta}_{vm}$ (i.e., $W^{\text{i}}_{vn}$); the fourth equation calculates the water outflow from reservoir $n$ during $L^{\Delta}_{vn}$ (i.e., $W^{\text{o}}_{vn}$), where $L^{\Delta}_{vu}$ denotes the time difference by which reservoir $v$ discharges later than reservoir $u$, as described in the fifth equation. 

\begin{figure}[tb]
	\centering
 \vspace{-0mm}
		\includegraphics[width=\columnwidth]{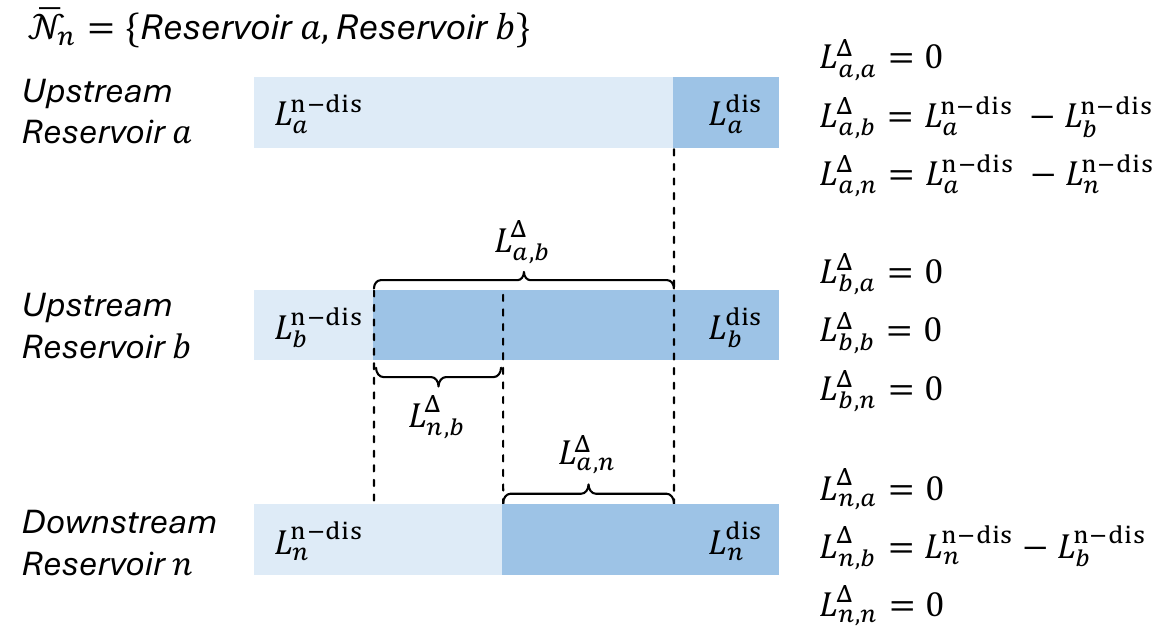}
  \vspace{-7mm}
	\caption{A demonstration of constraint \eqref{FModel:3}.}\label{Fig02}
	  \vspace{-3mm}
\end{figure}

Fig.~\ref{Fig02} further explains \eqref{FModel:3} using a three-reservoir example, where reservoirs $a$ and $b$ are direct upstream of reservoir $n$. The storage limit is checked when each of the reservoirs $a$, $b$, and $n$ begins discharging. When $a$ begins discharging, $b$ and $n$ have discharged for ($L^{\text{n-dis}}_{a}-L^{\text{n-dis}}_{b}$) and ($L^{\text{n-dis}}_{a}-L^{\text{n-dis}}_{n}$) weeks, respectively; when $b$ begins discharging, neither $a$ nor $n$ has discharged; when $n$ begins discharging, $a$ has not yet discharged, while $b$ has discharged for ($L^{\text{n-dis}}_{n}-L^{\text{n-dis}}_{b}$) weeks. 

Constraints \eqref{FModel:value}-\eqref{FModel:6} further enforce storage limits at the end of future period; constraint \eqref{FModel:7} describes the relationship between discharge rate and hydropower generation via a piece-wise linearized formulation; constraints \eqref{FModel:8} and \eqref{FModel:9} limit the discharge rate and hydropower generation level.

{\color{CBlue}
\textit{Remark:} Three specific settings of model \eqref{FModel} deserve further explanation and justification.}

{\color{CBlue}
\begin{itemize}
\item[\textit{i)}] \textit{The objective function \eqref{FModel:1} maximizes the total hydropower generation}. In existing medium-term hydropower planning works, objective functions broadly fall into three categories: maximizing hydropower generation \cite{type2_b}, maximizing operating profit based on forecasted market prices \cite{Arild1}, and minimizing operating costs \cite{Street}. This work adopts the first category to form the medium-term planning objective function \eqref{FModel:1}. This choice aligns with the fact that many hydropower operators lack reliable medium-term market price forecasting tools while the operating costs of their hydropower units are usually relatively low. Therefore, it is reasonable to prioritize total generation in their medium-term planning.

\item[\textit{ii)}] \textit{For each reservoir $n$, all non-discharge actions are aggregated into a single phase lasting $L^{\text{n-dis}}_{n}$ weeks, followed by one aggregated discharge phase lasting $L^{\text{dis}}_{n}$ weeks}. Due to the need to balance technical details and computational tractability in medium-term planning problems, aggregation has been adopted in seminal hydropower studies \cite{Norwegian_CHP} and remains a focus in more recent works \cite{Arild_aggregation}. Similarly, model \eqref{FModel}, whose full-model counterpart without aggregation is \eqref{EX} in Appendix~\ref{gap}, aggregates all discharge actions into a single phase. To explain this aggregation strategy, we consider a reservoir with the most efficient discharging rate of $X$ m\textsuperscript{3}/s for hydropower generation and operated over three sequential periods: period \textit{A} with 50 Mm\textsuperscript{3} discharge, period \textit{B} with 70 Mm\textsuperscript{3} charge and then 30 Mm\textsuperscript{3} discharge, and period \textit{C} with 15 Mm\textsuperscript{3} discharge. The maximum possible generation occurs when the reservoir discharges at the rate of $X$ m\textsuperscript{3}/s in all three periods. Because these discharge actions share the same rate, they can be aggregated into a single discharge phase, during which the total volume of (50 + 30 + 15) Mm\textsuperscript{3} is discharged at the rate of $X$ m\textsuperscript{3}/s. Moreover, because discharging all stored water by the end would achieve the maximum generation, we adopt a “charge-first, then-discharge” structure. Using the aggregation can free model \eqref{FModel} from relying on the time-step index. Thus, the model can scale to lengthy future periods (i.e., larger $L$) without inflating the model size. Meanwhile, constraint \eqref{FModel:3} preserves essential time-related details (e.g., discharge delays between connected reservoirs) so that the resulting discrepancies remain acceptable (about 2\% for the PGE case). Further discussion on the impact of aggregation is provided in Appendix~\ref{gap}.

\item[\textit{iii)}] \textit{Model \eqref{FModel} adopts the deterministic WI prediction} $\hat{W}^\text{fp}_{n}$.
There are two main reasons for adopting a deterministic prediction in model \eqref{FModel}. First, there is a trade-off between thoroughly representing prediction uncertainties and effectively deriving a straightforward future value function. Incorporating stochastic scenarios or uncertainty sets would significantly complicate this function, making it more difficult to derive and interpret, while keeping a deterministic form of model \eqref{FModel} can significantly ease the derivation process. Second, the derived future value function offers a strategic medium-term reference with balanced immediate hydropower generation and potential future gains to guide short-term operations. With this, model \eqref{FModel} and its resulting future value quantification \(F(V_n^{\text{cs},\theta})\) can be properly built based on a broad indication of future hydrological conditions (e.g., whether the period is wetter or drier). Thus, adopting a deterministic setting for model \eqref{FModel} can preserve tractability and interoperability in deriving the future value function without overly compromising planning effectiveness.
\end{itemize}
}

In model \eqref{FModel}, two issues need to be further tackled: deriving the WI predictions used in \eqref{FModel:value}-\eqref{FModel:5} and handling the bilinear terms in \eqref{FModel:1}, \eqref{FModel:3}, and \eqref{FModel:6}.

\subsubsection{Derive WI Predictions for Model \eqref{FModel}}\label{BMDN}
\begin{figure}[tb]
	\centering
		\includegraphics[width=\columnwidth]{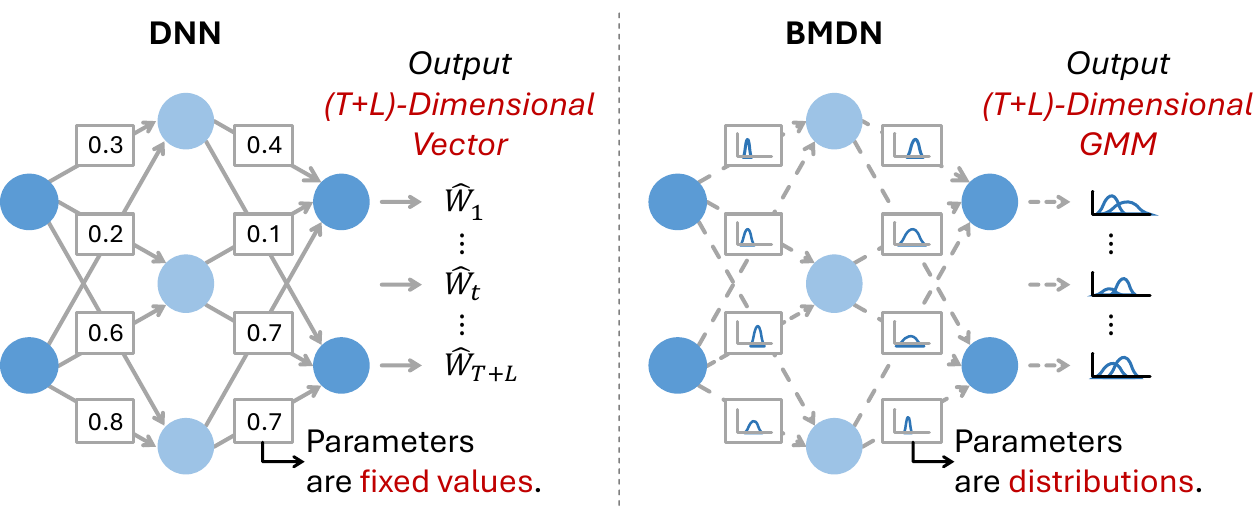}
  \vspace{-7mm}
	\caption{Comparison of DNN and BMDN.}\label{Fig03}
	\vspace{-3mm}
\end{figure}

Various data-driven predictors can be used to generate WI forecasts for model \eqref{FModel}.
{\color{CBlue}
This paper selects the Bayesian mixture density network (BMDN) \cite{BNN_1} because it offers high predictive accuracy and considerable flexibility for uncertainty modeling. Fig.~\ref{Fig03} compares BMDN with the conventional deep neural network (DNN), highlighting two key advantages:
\begin{itemize}
\item[\textit{i)}] \textit{Treatment of Network Weight Parameters:}
The conventional DNN idealistically assumes that the training dataset is perfect; therefore, its weight parameters are deterministic scalars. In contrast, the BMDN uses distribution-based weight parameters to account for the unavoidable imperfections of the dataset. As a result, BMDN often achieves higher predictive accuracy than DNN;

\item[\textit{ii)}] \textit{Form of Predictions:}
The DNN only provides point forecasts, which suffice the need of model \eqref{FModel} but are inadequate for the current-period model \eqref{General:2} that considers stochastic WIs. As compared, BMDN offers Gaussian mixture model (GMM)-based predictions that include both expected outcomes (as mean vectors) and potential variations (as covariance matrices). This versatile prediction enables the flexibility of adopting various optimization methods to manage uncertainties. For example, its affine invariance property can be leveraged to derive uncertainty quantiles for efficiently handling chance-constrained programming (CCP) models.
\end{itemize}
}

The WI predictions provided by BMDN can be expressed as \eqref{GMM}, a $(T+L)$-dimensional GMM corresponding to the $(T+L)$ weeks of the planning horizon. GMM \eqref{GMM} is a linear combination of $G$ Gaussian distribution components. For each component $g$, $\beta_{g}$ denotes its weight; $\phi(\cdot)$ and $\Phi(\cdot)$ are the probability density and cumulative distribution functions, respectively; $\boldsymbol{\mu}_{g}$ is the mean vector, playing the role similar to point predictions; $\boldsymbol{\Sigma}_{g}$ is the covariance matrix, informing the aleatoric uncertainty. In \eqref{GMM}, $\beta_{g}$, $\boldsymbol{\mu}_{g}$, and $\boldsymbol{\Sigma}_{g}$ are the prediction results, while $G$ is a pre-determined parameter.
\begin{flalign}\label{GMM}
\renewcommand{\arraystretch}{1.3}
\left\{ \begin{array}{ll}
\text{PDF}(\boldsymbol{\xi}) = {\sum\nolimits_{g = 1}^{G}\beta_{g} \times \phi(\boldsymbol{\xi}|\boldsymbol{\mu}_{g}, \boldsymbol{\Sigma}_{g});}\\
\text{CDF}(\boldsymbol{\xi}) = {\sum\nolimits_{g = 1}^{G}\beta_{g} \times \Phi(\boldsymbol{\xi});}\\
{\beta_{g} \geq 0, \sum\nolimits_{g=1}^{G} \beta_{g} = 1;}\\
\boldsymbol{\mu}_{g} \in \mathbb{R}^{(T+L)},
\boldsymbol{\Sigma}_{g} \in \mathbb{R}^{(T+L) \times (T+L)},g = 1,...,G;\\
\end{array}\right\}
\end{flalign}

It is noteworthy that GMM inherits the affine invariance of Gaussian distribution \cite{Wu_UC}. By leveraging this property, the prediction in \eqref{FModel:3}-\eqref{FModel:5} can be calculated following \eqref{setting}. 
\begin{equation}\label{setting}
\textstyle{\hat{W}_{n}^{\text{fp}} = \sum_{g = 1}^{G} \beta_{n,g} \sum_{k = T+1}^{T+L} \mu_{n,g,k}, \forall n;}
\end{equation}

\subsubsection{Convert Model \eqref{FModel} into Mixed-Integer Linear Programming}
Model \eqref{FModel} is a mixed-integer nonlinear programming model with bilinear terms of two continuous variables in \eqref{FModel:1}, \eqref{FModel:3}, and \eqref{FModel:6}. To this end, the binary expansion method \cite{binaryexpansion} is applied to represent $L_{n}^{\text{dis}}$ via the linear approximation \eqref{binaryExp} with auxiliary variables $\acute{L}_{nd}^{\text{dis}}$. Note that $\lfloor \cdot \rfloor$ is the floor function. {\color{CBlue}Parameter $\omega$ can be properly tuned to achieve the trade-off between approximation accuracy and computational complexity \cite{binaryexpansion, Sddip}.}
\begin{subequations}\label{binaryExp}
\begin{align}
&L_{n}^{\text{dis}}=\textstyle{L\times(1-\omega)\times\sum_{d=1}^{\lfloor \log_{2}\frac{1}{1-\omega} \rfloor + 1}2^{d-1} \acute{L}_{n,d}^{\text{dis}},}                          \mspace{-250mu}&\forall n;          \label{binaryExp:1}\\
&\textstyle{\acute{L}_{n,d}^{\text{dis}} \in\{0, 1\},}\mspace{-250mu}& \forall n, \forall d\in \{1,\cdots,\textstyle{\lfloor \log_{2}\frac{1}{1-\omega} \rfloor} + 1\};\label{binaryExp:2}
\end{align}
\end{subequations}

After using the binary expansion, bilinear terms in \eqref{FModel:1}, \eqref{FModel:3}, and \eqref{FModel:6} become products of binary and continuous variables that can be linearized via the big-M method. Thus, model \eqref{FModel} can be reformulated as a more tractable mixed-integer linear programming (MILP) problem whose compact form is \eqref{CModel}. Here, $\boldsymbol{c}$, $\boldsymbol{d}$, $\boldsymbol{A}$, $\boldsymbol{E}$, $\boldsymbol{b}$, and $\boldsymbol{F}$ are constant vectors and matrices determined via \eqref{FModel}, \eqref{setting}, and \eqref{binaryExp}.
\begin{subequations}\label{CModel}
\begin{align}
&{\max_{\boldsymbol{x}, \boldsymbol{y}}}\,\, \boldsymbol{c}^{\top}\boldsymbol{x} + \boldsymbol{d}^{\top}\boldsymbol{y}  \label{CModel:1} \\
&\text{s.t. }\boldsymbol{Ax}+\boldsymbol{Ey} \leq \boldsymbol{b}+\boldsymbol{F} \boldsymbol{V}^{\text{cs},\theta};   \label{CModel:2}\\
&\mspace{27mu}\boldsymbol{x} \in \mathbb{R}^{p}_{+},\, \boldsymbol{y} \in \{0, 1\}^{q};                            \label{CModel:3}
\end{align}
\end{subequations}

With specified parameter $\boldsymbol{V}^{\text{cs},\theta}$ and given WI predictions, the maximum possible hydropower generation over the future period can be determined by solving \eqref{CModel}. With this, the LMWV and future value can be formally defined.

\begin{figure}[tb]
	\centering
 \vspace{-0mm}
		\includegraphics[width=0.95\columnwidth]{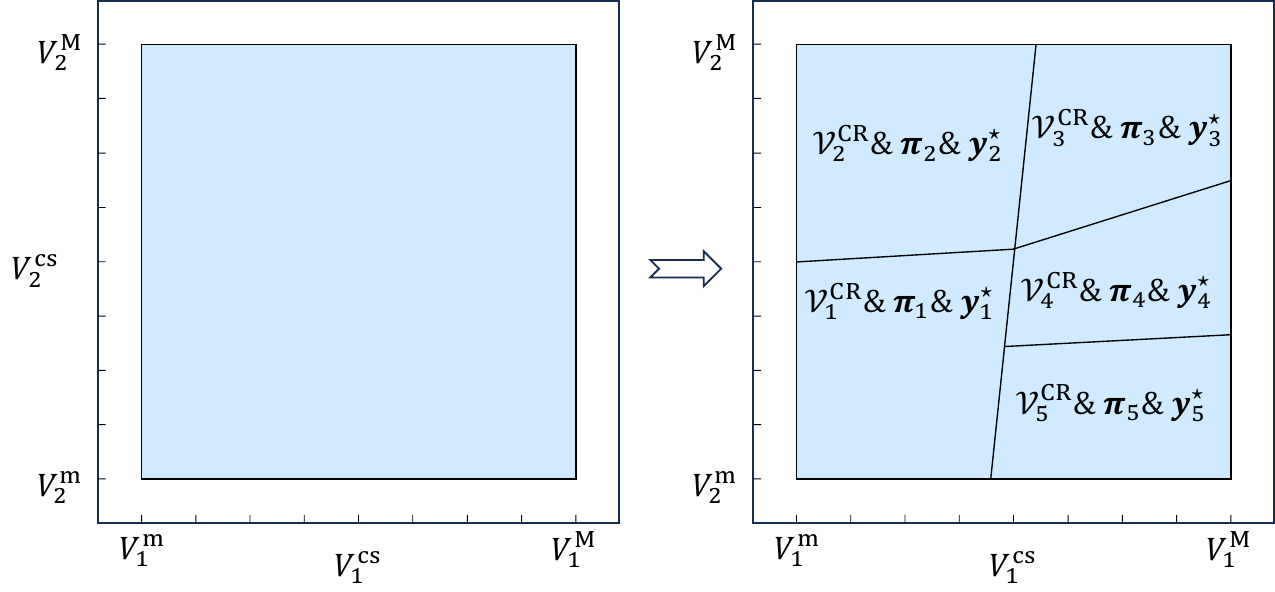}
  \vspace{-3mm}
	\caption{A partition-then-extract illustration with $N=\text{2}$ and $R=\text{5}$.}\label{Fig04}
	\vspace{-5mm}
\end{figure}

\begin{definition}\label{LMWV}
The LMWV of reservoir $n$ represents the increase in objective \eqref{CModel:1} caused by an additional 1 {\color{CBlue}Mm\textsuperscript{3}} of water in reservoir $n$. As model \eqref{CModel} will use all stored water by the end of the future period, the LMWV of reservoir $n$ corresponds to the dual variable of the $n^{\text{th}}$ constraint \eqref{FModel:value}.
\end{definition}

\begin{definition}\label{FV}
The \textit{future value} of carryover storage $\boldsymbol{V}^{\text{cs}}$ is defined as its exclusive contribution to the objective \eqref{CModel:1}, calculated by $\sum_{n=1}^{N} \text{LMWV}_{n} \times (V^{\text{cs}}_{n} - V^{\text{m}}_{n})$.
\end{definition}

\vspace{-0mm}
\subsection{Use Partition-then-Extract Algorithm to Calculate LMWVs}\label{Step2}
\vspace{-0mm}
Given Definitions~\ref{LMWV} and \ref{FV}, it is clear that identifying the LMWV is key to quantifying the future value.
However, identifying the LMWV is intractable as it depends on both the active set of constraints and the binary variable solutions in model \eqref{CModel}, both of which vary with specified $\boldsymbol{V}^{\text{cs},\theta}$ values. Therefore, this paper draws the idea of multi-parametric mixed-integer linear programming (mp-MILP) \cite{mpMILP_app1} and treats $\boldsymbol{V}^{\text{cs},\theta}$ as a parameter vector. Consequently, a partition-then-extract algorithm is applied to \eqref{CModel} for calculating LMWVs.

Fig.~\ref{Fig04} illustrates the concept behind the partition-then-extract algorithm: the entire feasible space of $\boldsymbol{V}^{\text{cs},\theta}$ (denoted as $\mathcal{V}^{\text{E}} = \{V^{\text{cs},\theta}_{n}| V_{n}^{\text{m}} \leq V^{\text{cs},\theta}_{n} \leq V_{n}^{\text{M}}, \forall n \in \mathcal{N} \}$ and represented by the blue area) is partitioned into $R$ critical regions (CRs) $\mathcal{V}_{r}^{\text{CR}}$ for $r=1,...,R$. Each $\mathcal{V}_{r}^{\text{CR}}$ is associated with 
a unique optimal binary solution (denoted as $\boldsymbol{y}^{\star}_r$) and a unique LMWV vector (denoted as $\boldsymbol{\pi}_{r}$), as explained in Definition~\ref{CR}.

\begin{definition}\label{CR}
In mp-MILP, a CR is a polyhedral subspace of the parameter space where the active set of constraints and the values of the binary variables \textit{remain unchanged} \cite{mpMILP_app2}. The parameter space $\mathcal{V}^{\text{E}}$ can be partitioned into $R$ critical regions satisfying exhaustiveness ($\cup_{r=\text{1}}^{R} \mathcal{V}_{r}^{\text{CR}} = \mathcal{V}^{\text{E}}$) and disjointness ($\mathcal{V}_{r}^{\text{CR}} \cap \mathcal{V}_{h}^{\text{CR}} = \emptyset $ for $r \neq h$).
\end{definition}

The partition-then-extract algorithm, as detailed in Algorithm~\ref{alg0}, involves three main steps:
(i) solving a multi-parametric linear programming (mp-LP) problem \eqref{mpLP} with a given binary solution to partition a certain carryover storage space into multiple CRs \cite{mpLP_app1};
(ii) solving an MILP problem \eqref{MILPsub} to explore better binary solutions for a given CR;
and (iii) solving a dual problem \eqref{Dual} to calculate the LMWV vector for each CR.
The algorithm iterates between steps (i) and (ii) to explore all CRs of the entire parameter space $\mathcal{V}^{\text{E}}$, and then uses step (iii) to calculate LMWVs for each of the final CRs.

\begin{algorithm}[tb]
\caption{Partition-then-Extract Algorithm}\label{alg0}
	\SetAlgoLined
	\KwIn{

    $\mspace{6mu}\mathcal{V}^{\text{E}}$: the entire parameter space, $\bar{\boldsymbol{y}}^{\text{ini}}$: an initial binary solution to \eqref{CModel}.

	}
         
	\KwOut{
	
    $\mspace{6mu}\mathcal{L} = \{[\mathcal{V}_{1}^{\text{CR}}, \boldsymbol{\pi}_{1}],..., [\mathcal{V}_{R}^{\text{CR}}, \boldsymbol{\pi}_{R}]\}$: set of CR and LMWV tuples.

	}
	 
\textbf{Initialization:}

 $\mspace{6mu}$Initialize the ordered set $\mathcal{D} = \{[\mathcal{V}^{\text{E}}, -\infty, \bar{\boldsymbol{y}}^{\text{ini}}]\}$.  The three elements in each tuple of $\mathcal{D}$ are a CR, its best lower bound (i.e., IPLB), and a set of binary solutions that have been explored for this CR. Initialize an empty set $\mathcal{L}$ to record final CR and LMWV tuples.

\While{$\mathcal{D} \neq \emptyset$}{
$\mspace{23mu}$Select the first tuple in $\mathcal{D}$, denoted as $[\mathcal{\hat{V}}^{\text{CR}}, \hat{z}(\boldsymbol{{V}}^{\text{cs},\theta})^{\text{LB}}, \mathcal{\hat{B}}]$, and remove this tuple from $\mathcal{D}$\;

$\mspace{23mu}$Solve \eqref{MILPsub} with respect to $[\mathcal{\hat{V}}^{\text{CR}}, \hat{z}(\boldsymbol{{V}}^{\text{cs},\theta})^{\text{LB}}, \mathcal{\hat{B}}]$\;

\uIf{\eqref{MILPsub} has an optimal solution (better binary solutions exist)}{
	$\mspace{23mu}$Add the optimal binary solution of \eqref{MILPsub}, denoted as $\boldsymbol{y}^{\star}$, to the end of ordered set $\mathcal{\hat{B}}$\;
	
	$\mspace{23mu}$ Fix $\bar{\boldsymbol{y}}$ of \eqref{mpLP} as $ \boldsymbol{y}^{\star}$; Fix $\mathcal{V}^{\text{P}}$ of \eqref{mpLP:C} as $\mathcal{\hat{V}}^{\text{CR}}$\;
	
	\textbf{Partition} $\mathcal{\hat{V}}^{\text{CR}}$\textbf{:}
 
	$\mspace{23mu}$Solve mp-LP \eqref{mpLP}, which partitions $\mathcal{\hat{V}}^{\text{CR}}$\ into $C$ CRs denoted as $\mathcal{\hat{V}}^{\text{CR}}_1,...,\mathcal{\hat{V}}^{\text{CR}}_C$, together with their incumbent parametric lower bounds $\hat{z}(\boldsymbol{V}^{\text{cs}})_{1}^{\text{LB}},..., \hat{z}(\boldsymbol{V}^{\text{cs}})_{C}^{\text{LB}}$;
	
	$\mspace{23mu}$Create $C$ ordered sets $\mathcal{\hat{B}}_{1},...,\mathcal{\hat{B}}_{C}$, and initialize each of them as $\mathcal{\hat{B}}$\;

	$\mspace{23mu}$Add the $C$ tuples $[\mathcal{\hat{V}}^{\text{CR}}_{h}, \hat{z}(\boldsymbol{V}^{\text{cs}})_{h}^{\text{LB}}, {\hat{B}}_{h}]$ for $h=\{1,...,C\}$ to the end of ordered set $\mathcal{D}$\;}

	\ElseIf{\eqref{MILPsub} is infeasible (no better binary solution exists)}{
       Identify $\mathcal{\hat{V}}^{\text{CR}}$ as a final CR, re-denoted as $\mathcal{V}^{\text{CR}}_{r}$\;
        
      Fix ${\boldsymbol{y}}_{r}^{\star}$ of \eqref{Dual:A} as the last element in ordered set $\mathcal{\hat{B}}$\;

      Fix $\hat{\boldsymbol{V}}^{\text{cs}}_{r}$ of \eqref{Dual:A} as an arbitrary point in $\mathcal{V}^{\text{CR}}_{r}$\;
  
        \textbf{Calculate LMWVs:}
        
        $\mspace{23mu}$Solve the LP problem \eqref{Dual} to calculate the LMWV vector for $\mathcal{V}^{\text{CR}}_{r}$, denoted as $\boldsymbol{\pi}_{r}$\;

        $\mspace{23mu}$Create a new tuple [$\mathcal{V}^{\text{CR}}_{r}$, $\boldsymbol{\pi}_{r}$] and add it to set $\mathcal{L}$;}
  }
\end{algorithm}

\subsubsection{Partition A Given Carryover Storage Space into Multiple CRs}\label{problem1}
The mp-LP problem with a given binary variable solution $\bar{\boldsymbol{y}}$ is formulated as \eqref{mpLP}. Note that $\bar{\boldsymbol{y}}$ is initiated via a feasible binary solution and then is iteratively improved via solving \eqref{MILPsub} in Section \ref{problem2}. 
\begin{subequations}\label{mpLP}
\begin{align}
&J(\boldsymbol{V}^{\text{cs},\theta}) =
        {\max_{\boldsymbol{x}}\boldsymbol{c}^{\top}\boldsymbol{x} + \boldsymbol{d}^{\top} \bar{\boldsymbol{y}}} \label{mpLP:A}\\
&\text{s.t. }\boldsymbol{Ax}+\boldsymbol{E}\bar{\boldsymbol{y}} \leq \boldsymbol{b} + \boldsymbol{F} \boldsymbol{V}^{\text{cs},\theta};\label{mpLP:B}\\
&\mspace{26mu}\boldsymbol{x} \in \mathbb{R}_{+}^{p};\,\boldsymbol{V}^{\text{cs},\theta} \in \mathcal{V}^{\text{P}}; \label{mpLP:C}
\end{align}
\end{subequations}

Solving mp-LP \eqref{mpLP} will partition space $\mathcal{V}^{\text{P}}$ into $C$ CRs\footnote{{\color{CBlue}This paper uses the comparison procedure from \cite{Compare} to handle potential degeneracy when solving mp-LP \eqref{mpLP}. When finalizing each CR, redundant constraints are further removed according to the approach described in \cite{book}.}}, denoted as $\mathcal{\hat{V}}_{h}^{\text{CR}}$ for $h=\{1,...,C\}$. Each $\mathcal{\hat{V}}_{h}^{\text{CR}}$ is a polytope described via $O_{h}$ linear inequalities \eqref{Polytope}, where $\boldsymbol{e}$ and $f$ are constant vectors and scalars.
\begin{equation}\label{Polytope}
\mathcal{\hat{V}}_{h}^{\text{CR}}=
\{\boldsymbol{V}^{\text{cs},\theta}\,|\,\boldsymbol{e}_{h,o}^{\top}\boldsymbol{V}^{\text{cs},\theta}+f_{h,o} \leq 0,\, o=1,...,O_{h}\}
\end{equation}

Solving \eqref{mpLP} also provides incumbent parametric lower bounds (IPLB), $\hat{z}(\boldsymbol{V}^{\text{cs},\theta})_{h}^{\text{LB}}$, which will be used in model \eqref{MILPsub} of Section~\ref{problem2} to explore better binary solutions for $\mathcal{\hat{V}}_{h}^{\text{CR}}$.

\subsubsection{Explore Better Binary Solutions for Each Given CR}\label{problem2}
For a given CR $\mathcal{\hat{V}}^{\text{CR}}$, together with its IPLB $\hat{z}(\boldsymbol{V}^{\text{cs},\theta})^{\text{LB}}$ and an ordered set of binary solutions $\mathcal{\hat{B}}$ that have been obtained, the exploration problem is constructed as \eqref{MILPsub}. It uses a bounding cut \eqref{MILPsub:C} and an integer cut \eqref{MILPsub:D} to search for better binary solutions. In \eqref{MILPsub:C}, threshold $\varrho$ is a sufficiently small positive constant. In \eqref{MILPsub:D}, sets $\mathbb{I}_{b} = (j|\bar{y}_{b,j}=1)$ and $\mathbb{O}_{b} = (j|\bar{y}_{b,j}=0)$ are built on each of explored solutions $b \in \mathcal{\hat{B}}$. If \eqref{MILPsub} is feasible, it delivers a better binary solution for $\mathcal{\hat{V}}^{\text{CR}}$, which is then added to $\mathcal{\hat{B}}$ and fed back into \eqref{mpLP} for further partitioning $\mathcal{\hat{V}}^{\text{CR}}$; otherwise, $\mathcal{\hat{V}}^{\text{CR}}$ is one of the final $R$ CRs (denoted as $\mathcal{V}^{\text{CR}}_{r}$) that does not require further partition, and its best binary solution, denoted as $\boldsymbol{y}_{r}^{\star}$, is the last record in $\mathcal{\hat{B}}$.
\begin{subequations}\label{MILPsub}
\begin{align}
{\max_{\boldsymbol{x},\boldsymbol{y},\boldsymbol{V}^{\text{cs}}}}
&\textstyle{\,\,\boldsymbol{c}^{\top} \boldsymbol{x} + \boldsymbol{d}^{\top} \boldsymbol{y}}   \mspace{-30mu}&\label{MILPsub:A}\\
&\mspace{-80mu}\text{s.t. }\boldsymbol{Ax}+\boldsymbol{Ey} \leq \boldsymbol{b} + \boldsymbol{F} \boldsymbol{V}^{\text{cs}},\,\boldsymbol{x} \in \mathbb{R}_{+}^{p},\, \boldsymbol{y} \in \{0, 1\}^{q};                                     \mspace{-30mu}&\label{MILPsub:B}\\
&\mspace{-52mu}\boldsymbol{c}^{\top}\boldsymbol{x} + \boldsymbol{d}^{\top} \boldsymbol{y} \geq \hat{z}(\boldsymbol{V}^{\text{cs},\theta})^{\text{LB}} + \varrho;                                                                                   \mspace{-30mu}&\label{MILPsub:C}\\
&\mspace{-52mu}\textstyle{\sum\nolimits_{j \in \mathbb{I}_{b}}y_{b,j} - \sum\nolimits_{j \in \mathbb{O}_{b}} y_{b,j} \leq |\mathbb{I}_{b}| - 1, \forall b \in \mathcal{\tilde{B}};}                                                                \mspace{-30mu}&\label{MILPsub:D}\\
&\mspace{-52mu}\boldsymbol{V}^{\text{cs}} \in \mathcal{\hat{V}}^{\text{CR}};               \mspace{-30mu}&\label{MILPsub:E}
\end{align}
\end{subequations}

\subsubsection{Calculate LMWVs for a Final CR}\label{problem3}
For each $\mathcal{V}^{\text{CR}}_{r}$, model \eqref{Dual} is solved to calculate its LMWV vector $\boldsymbol{\pi}_{r}$.
\begin{subequations}\label{Dual}
\begin{align}
&{\min_{\boldsymbol{\varphi}} (\boldsymbol{b} + \boldsymbol{F} \hat{\boldsymbol{V}}^{\text{cs}}_{r} - \boldsymbol{E} \boldsymbol{y}_{r}^{\star})^{\top}\boldsymbol{\varphi}}\label{Dual:A}\\
&\text{s.t. } \boldsymbol{A}^{\top}\boldsymbol{\varphi} \geq \boldsymbol{c},\,\boldsymbol{\varphi} \in \mathbb{R}_{+}^{d}          ;\label{Dual:D}
\end{align}
\end{subequations}

Model \eqref{Dual} is an LP problem built based on \eqref{CModel} via three steps:
\textit{i)} fix binary variables $\boldsymbol{y}$ in \eqref{CModel} as $\boldsymbol{y}_{r}^{\star}$ (obtained in Section \ref{problem2} by solving \eqref{MILPsub}) to convert \eqref{CModel} into an LP problem with parameter $\boldsymbol{V}^{\text{cs},\theta}$;
\textit{ii)} set $\boldsymbol{V}^{\text{cs},\theta}$ as an arbitrary point (denoted as $\hat{\boldsymbol{V}}^{\text{cs}}_{r}$) within $\mathcal{V}^{\text{CR}}_{r}$, resulting in a standard LP model; and 
\textit{iii)} dualize the LP from step \textit{ii)} to build model \eqref{Dual}.

Based on Definitions~\ref{LMWV} and \ref{CR}, LMWVs $\pi_{rn}$ of reservoir $n$ under carryover storage levels within $\mathcal{V}^{\text{CR}}_{r}$ are the optimal $\boldsymbol{\varphi}$ corresponding to \eqref{FModel:value} for reservoir $n$.

\vspace{-0mm}
\subsection{Construct Analytical Quantification Rules of Future Value}\label{Step3}
\vspace{-0mm}
The partition-then-extract algorithm ultimately yields $R$ LMWV vectors $\boldsymbol{\pi}_{1},...,\boldsymbol{\pi}_{R}$ corresponding to the $R$ final CRs $\mathcal{V}_{1}^{\text{CR}},...,\mathcal{V}_{R}^{\text{CR}}$. According to Definition~\ref{FV}, the future value of carryover storage $\boldsymbol{V}^{\text{cs}}$ under the predicted WI $\hat{\boldsymbol{W}}^{\text{fp}}$ can be calculated via the following ``if-then'' rules \eqref{Ifthen}.
\begin{align}\label{Ifthen}
&F(\boldsymbol{V}^{\text{cs}}) \mspace{-3mu} = \mspace{-3mu} \left\{ \mspace{-10mu} \begin{array}{cl}
\sum_{n=1}^N \pi_{1,n}(V^{\text{cs}}_{n} - V^{\text{m}}_{n})&\mspace{-10mu} \text{if } \boldsymbol{V}^{\text{cs}} \mspace{-3mu} \in \mspace{-3mu} \mathcal{V}_{1}^{\text{CR}};\\
                          \vdots                            &\\
\sum_{n=1}^N \pi_{R,n}(V^{\text{cs}}_{n} - V^{\text{m}}_{n})&\mspace{-10mu} \text{if } \boldsymbol{V}^{\text{cs}} \mspace{-3mu} \in \mspace{-3mu}
 \mathcal{V}_{R}^{\text{CR}};\\
\end{array}\right.\mspace{-20mu}&
\end{align}

The ``if-then'' rules \eqref{Ifthen} offers three advantages:
\begin{itemize}
\item[\textit{i)}]
\textit{Interpretable.}
It can be interpreted as that if the carryover storage $\boldsymbol{V}^{\text{cs}}=[V^{\text{cs}}_{1},...,V^{\text{cs}}_{n},...,V^{\text{cs}}_{N}]^{\top}$ falls within region $\mathcal{V}_{r}^{\text{CR}}$, then the LMWV vector will be $\boldsymbol{\pi}_{r}=[\pi_{r,1},...,\pi_{r,n},...,\pi_{r,N}]^{\top}$ and the corresponding future value can be quantified as $\sum_{n=1}^N \pi_{r,n}(V^{\text{cs}}_{n} - V^{\text{m}}_{n})$;

\item[\textit{ii)}]
\textit{Hydrologically adaptive.}
The calculation of LMWVs explicitly considers the WI predictions, thus being adaptive to future hydrological conditions. Specifically, LMWVs are low/high if future WIs are sufficient/insufficient. This fact will be further illustrated in the case studies in Section~\ref{PGEInsight};

\item[\textit{iii)}]
\textit{Easy-to-use.} By introducing $R$ binary variables $Z_{r}$ and a sufficiently large constant $M$, rule \eqref{Ifthen} can be converted into linear constraints \eqref{Easytouse}, which can be directly embedded into the medium-term planning model \eqref{General} to analytically represent the abstract future value function $F(\cdot)$ in \eqref{General:1}.
\begin{subequations}\label{Easytouse}
\begin{align}
           & \textstyle{F(\boldsymbol{V}^{\text{cs}})=\sum_{r = 1}^{R}Z_{r}\sum_{n=1}^{N} \pi_{r,n}(V^{\text{cs}}_{n} - V^{\text{m}}_{n});}                                        \label{Easytouse:1}\\
           &\textstyle{\sum_{r = 1}^{R} Z_{r} = 1; Z_{r} \in \{0, 1\}, \forall r};\label{Easytouse:2}\\
           &\textstyle{\left\{ \begin{array}{ll}
            \boldsymbol{e}_{r,1}^{\top}     \boldsymbol{V}^{\text{cs}} + f_{r,1}     \leq M(1-Z_{r}),  &\forall r;\\
            \mspace{30mu}                   \vdots                                       \\
            \boldsymbol{e}_{r,O_{r}}^{\top} \boldsymbol{V}^{\text{cs}} + f_{r,O_{r}} \leq M(1-Z_{r}),  &\forall r;
                       \end{array}\right.} \label{Easytouse:3}
\end{align}
\end{subequations}
\end{itemize}

\vspace{-0mm}
\section{Case Studies on the PGE System}\label{Sec4}
\vspace{-0mm}
\subsection{Pelton-Round Butte CHP of PGE: A Glance}
\vspace{-0mm}
\begin{figure}[tb]
	\centering
 \vspace{-0mm}
		\includegraphics[width=\columnwidth]{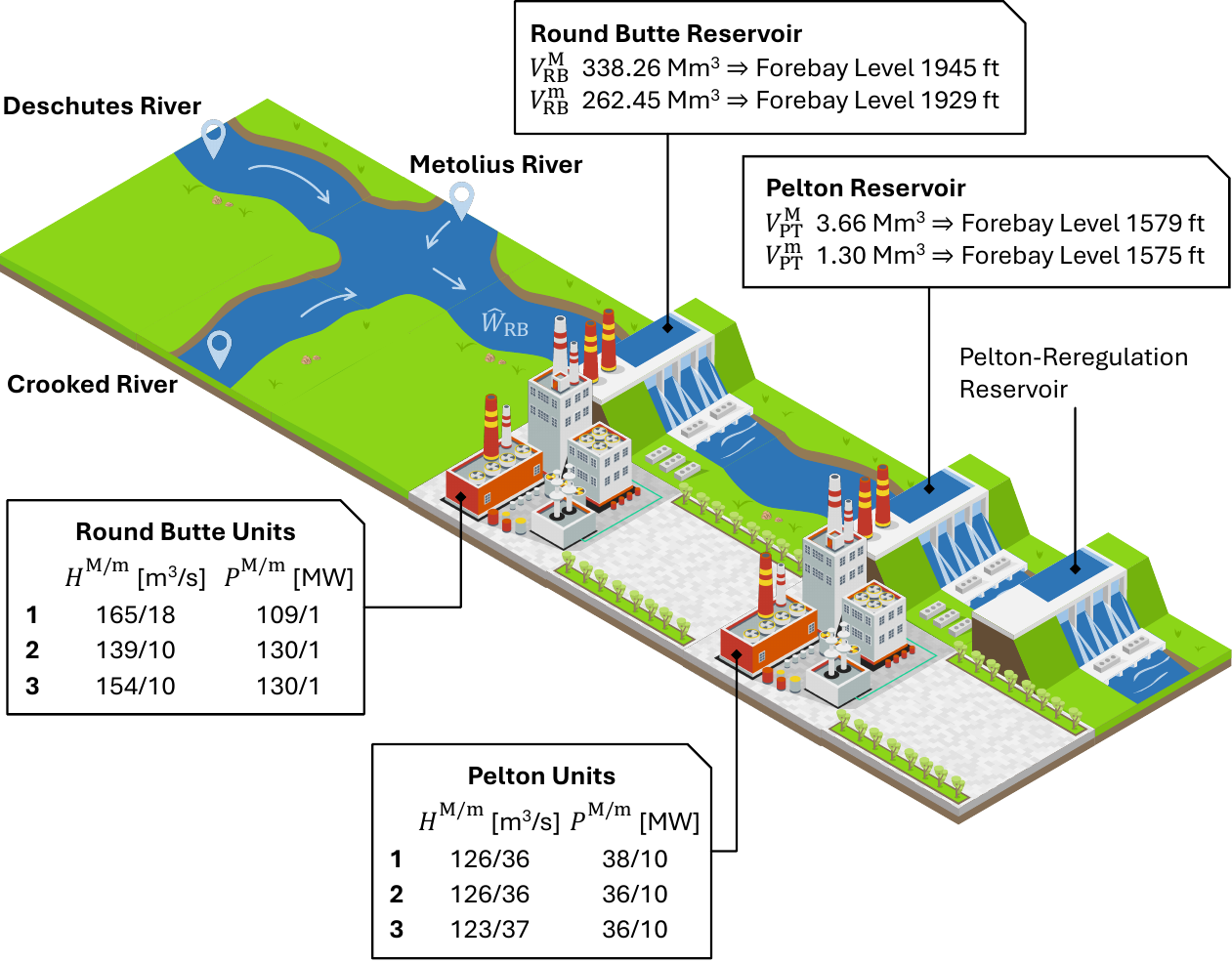}
  \vspace{-3mm}
	\caption{Illustration of PGE's Pelton-Round Butte CHP.}\label{Fig05}
	\vspace{-0mm}
\end{figure}

PGE is an active participant in the Western U.S. electricity grid. This section uses one of PGE's CHPs to illustrate the effectiveness of the proposed study. As shown in Fig.~\ref{Fig05}, the selected CHP comprises three reservoirs, including the upstream Round Butte (RB), the middle Pelton (PT), and the downstream PT-Reregulation. The system receives natural WIs through the RB reservoir, mainly from three upstream branches: Crooked River, Deschutes River, and Metolius River.

According to the PGE operation manual, the PT-Reregulation reservoir was originally designed to maintain the natural WI of the Deschutes River rather than generate hydropower. Consequently, only the RB and PT reservoirs are included in the scheduling process. Each of these two reservoirs contains a powerhouse with three hydropower units. The water-to-power-conversion curves $\mathcal{P}^{\text{RtP}}(\cdot)$ for these six units are quadratic and monotonically increasing, which are piecewise linearized to form \eqref{FModel:7}. {\color{CBlue}Lastly, it is important to note that because PGE limits the forebay levels within a narrow range, the water head of the PT-RB CHP remains relatively stable, allowing PGE to use the same water-to-power-conversion curves throughout the season.}

\vspace{-0mm}
\subsection{Methods to be Compared}
\vspace{-0mm}
To evaluate the effectiveness of the proposed future value quantification framework, the ``if-then'' rules \eqref{Ifthen} in the form of linear constraints \eqref{Easytouse} are embedded in medium-term CHP planning models to calculate carryover storage for CHPs and assess its effects on immediate benefit and future value. Specifically, the following three medium-term planning models are compared via numerical case studies on the PT-RB CHP:
\begin{itemize}
\item
{\color{CBlue}
\textbf{CCP-BMDN:} This method solves a concrete version of model \eqref{General} to determine carryover storage, as detailed in Appendix \ref{DetailedModel1}. In the current-period part \eqref{General:2}, WI uncertainties are managed via CCP\footnote{{\color{CBlue}CCP is selected due to its great compatibility with GMM-based WI forecasts \cite{Wu_UC}, which can significantly streamline the simulation process.}} with GMM-based predictions (from BMDN presented in Section \ref{BMDN}). The future value function \(F(\cdot)\) uses the one derived by the presented quantification framework;
}

\item
\textbf{DET-EF:} This method, as detailed in Appendix~\ref{DetailedModel2}, solves a concrete counterpart of model \eqref{General} using error-free WI point predictions from PGE's dataset. These perfect predictions are applied to the current-period formulation \eqref{General:2} and to the presented quantification framework. Essentially, DET-EF is an ideal deterministic variant of CCP-BMDN, serving as a benchmark to show the impact of prediction quality on the quantification framework;

\item
\textbf{PGE-P:} This represents PGE's current practice that follows an experience-based and seasonally adaptive policy. Specifically, RB and PT reservoirs are heavily discharged from November to January (wetter seasons) and then gradually refilled from February to April (transition seasons) to ensure sufficient storage by May, preparing for drier seasons (May to October). PGE may adjust this policy using rule-of-thumb approaches and WI predictions from the Northwest River Forecast Center. The results are based on PGE's historical operation records.

\end{itemize}

\begin{figure}[tb]
	\centering
		\includegraphics[width=\columnwidth]{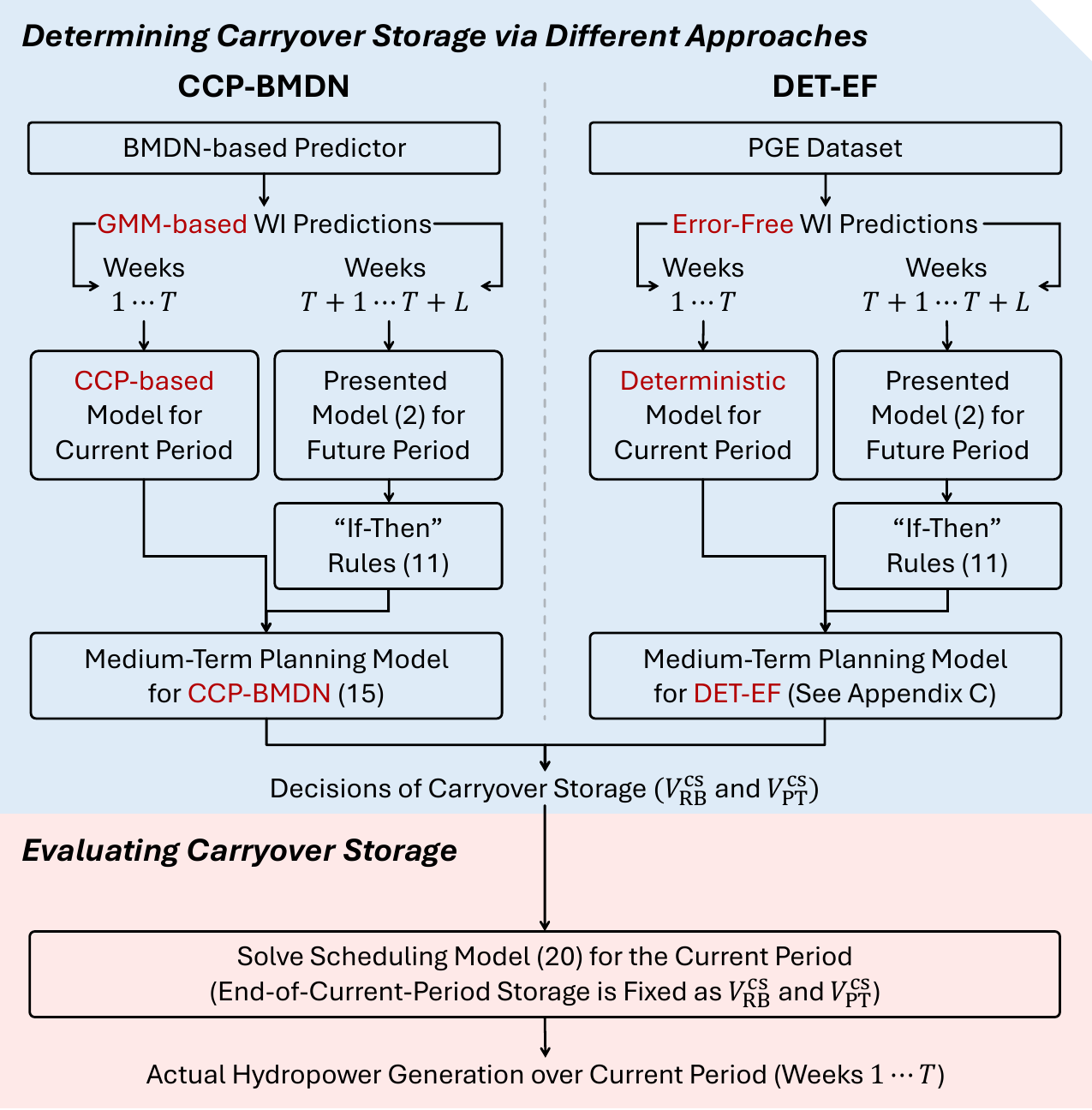}
  \vspace{-7mm}
	{\color{CBlue}\caption{Flowchart of determining and evaluating the carryover storage.}\label{Fig06}}
 \vspace{-0mm}
\end{figure}

{\color{CBlue}

Fig.~\ref{Fig06} illustrates the process of determining and evaluating carryover storage via CCP-BMDN and DET-EF. For each medium-term planning simulation run that covers a time span of $T$-week current period and $L$-week future period, LMWVs are calculated via the quantification framework \eqref{FModel} covering the $L$-week future period, which are then used to build future value function $F(\cdot)$ for CCP-BMDN and DET-EF to calculate the target carryover storage level \({V}^{\text{cs}}_{n}\) at the end of week $T$. After that, the short-term scheduling model \eqref{ST} covering the \(T\)-week current period is executed, using the target carryover storage level \({V}^{\text{cs}}_{n}\) as the right-hand-side parameter in constraint \eqref{ST:target} and incorporating the realized WI \(\tilde{W}_{nt}\). The hydropower generation derived by the objective function \eqref{ST:1} represents the immediate benefit induced by the target carryover storage \({V}^{\text{cs}}_{n}\), quantifying its effects in guiding short-term operations. The process then advances in an increment of $T$ weeks to the next medium-term planning simulation run.

As the PT–RB CHP can be fully drained down to its storage lower limit or recharged to its storage upper limit within one month, its medium-term planning horizon is typically set as 8 to 16 weeks in practice. Accordingly, this paper sets \(T = \text{4}\) to allow the simulation to proceed on a monthly basis. A sensitivity study on the future period length \(L\) ranging from \(L = \text{4}\) to \(L = \text{12}\) is conducted. Consequently, the BMDN produces \((\text{4} + L)\)-dimensional WI predictions, with each dimension corresponding to a weekly WI volume.
}

\vspace{-0mm}
\subsection{Experimental Settings}
\vspace{-0mm}

LP and MILP problems are solved via Gurobi 10.0, while mp-LP problems \eqref{mpLP} are solved following \cite{mpLP_app2}. All cases are conducted on a 2.4GHz PC.
BMDN is executed on TensorFlow 2.13. In each prediction run, BMDN is sampled $K$ times, generating $K$ GMM-based WI predictions \eqref{GMM}; their standard deviation vectors $\boldsymbol{\sigma}$ are calculated, and the GMM with the smallest $\lVert \boldsymbol{\sigma} \rVert_{1}$ is selected as the final WI predictions. Other details are available in our BMDN implementation codes \cite{BMDNCode}.

A PGE dataset spanning from 01/01/2010 to 12/31/2018 is used. It consists of \textit{i)} temperature and discharge rates of the three upstream rivers (input features of BMDN), \textit{ii)} El Niño indicators (input features of BMDN), \textit{iii)} actual WI realizations (output labels of BMDN), and \textit{iv)} the actual hydropower generation of the CHP. The data from 01/01/2010 to 12/31/2017 are used for training and validating the BMDN, while the remaining 2018 data are for out-of-sample testing.

\begin{table}[tb]
	\caption{Hyperparameter Setting Based on Cross Validation}\label{Tab01}
 \vspace{-0mm}
	\centering
	\footnotesize
\begin{tabular}{lcccc}
\toprule
                                    &DNN       &BMDN    \\
\midrule
Activation Type                     &sigmoid   &sigmoid \\
Loss Function                       &MSE       &Negative Log-Likelihood\\
Structure of Hidden Layers          &[4, 4]    &[4, 4]       \\
Number of Gaussian Components       &-         &2\\
\bottomrule
\end{tabular}
 \vspace{-0mm}
\end{table}

\begin{table}[tb]
	\caption{Statistical Comparison of Water Inflow Predictions}\label{Tab02}
  \vspace{-0mm}
	\centering
	\footnotesize
\begin{tabular}{lcccc}
\toprule
Predictor    &MAE/{\color{CBlue}Mm\textsuperscript{3}}&MAPE  &RMSE/{\color{CBlue}Mm\textsuperscript{3}}        &NSE  \\
\midrule
DNN            &{\color{CBlue}7.06}       &11.17\%    &{\color{CBlue}8.19}       &0.18   \\
BMDN-2         &{\color{CBlue}8.29}       &18.74\%    &{\color{CBlue}9.44}       &0.07   \\
BMDN-20        &{\color{CBlue}4.67}       &7.51\%     &{\color{CBlue}5.77}       &0.37   \\
\bottomrule
\end{tabular}
\vspace{-0mm}
\end{table}

\vspace{-0mm}
\subsection{Water Inflow Predictions}\label{WIPredictions}
\vspace{-0mm}
As WI predictions are needed for both the quantification framework (i.e., $\hat{\boldsymbol{W}}^{\text{fp}}$) and the CCP-BMDN model (for deriving chance constraints), this subsection first analyzes WI predictions provided by BMDN as presented in Section~\ref{BMDN}, using DNN  as a benchmark. Table \ref{Tab01} lists the major hyperparameters of BMDN and DNN. This subsection sets $L = \text{4}$.

Table~\ref{Tab02} lists the prediction performance assessed by four metrics: mean absolute error (MAE), mean absolute percentage error (MAPE), root mean square error (RMSE), and Nash–Sutcliffe model efficiency (NSE). BMDNs with different $K$ settings are tested, and BMDN-2 and BMDN-20 are reported in Table~\ref{Tab02}, respectively referring to BMDNs with $K =$ 2 and 20. Across all four metrics, BMDN-2 performs worse than DNN, but BMDN-20 outperforms DNN. Indeed, our extensive experiments indicate that the overall performance of BMDN gradually surpasses DNN once $K$ exceeds 5. This means that the way BMDN handles epistemic uncertainty\textemdash sampling from its distribution-based network and selecting the result with the smallest $\lVert \boldsymbol{\sigma} \rVert_{1}$ (i.e., the most confident)\textemdash can effectively enhance its prediction. MAE and MAPE results show that both BMDN-20 and DNN can offer good accuracy, while BMDN-20 is slightly better. Moreover, the RMSE results indicate that larger deviations between predictions and WI realizations occur less frequently in BMDN-20, meaning that BMDN-20 offers better stability. Finally, the NSE results indicate that BMDN-20 exhibits better predictive skills. Hereafter, BMDN-20 is used by default.

\begin{figure}[tb]
	\centering
 \vspace{-0mm}
		\includegraphics[width=\columnwidth]{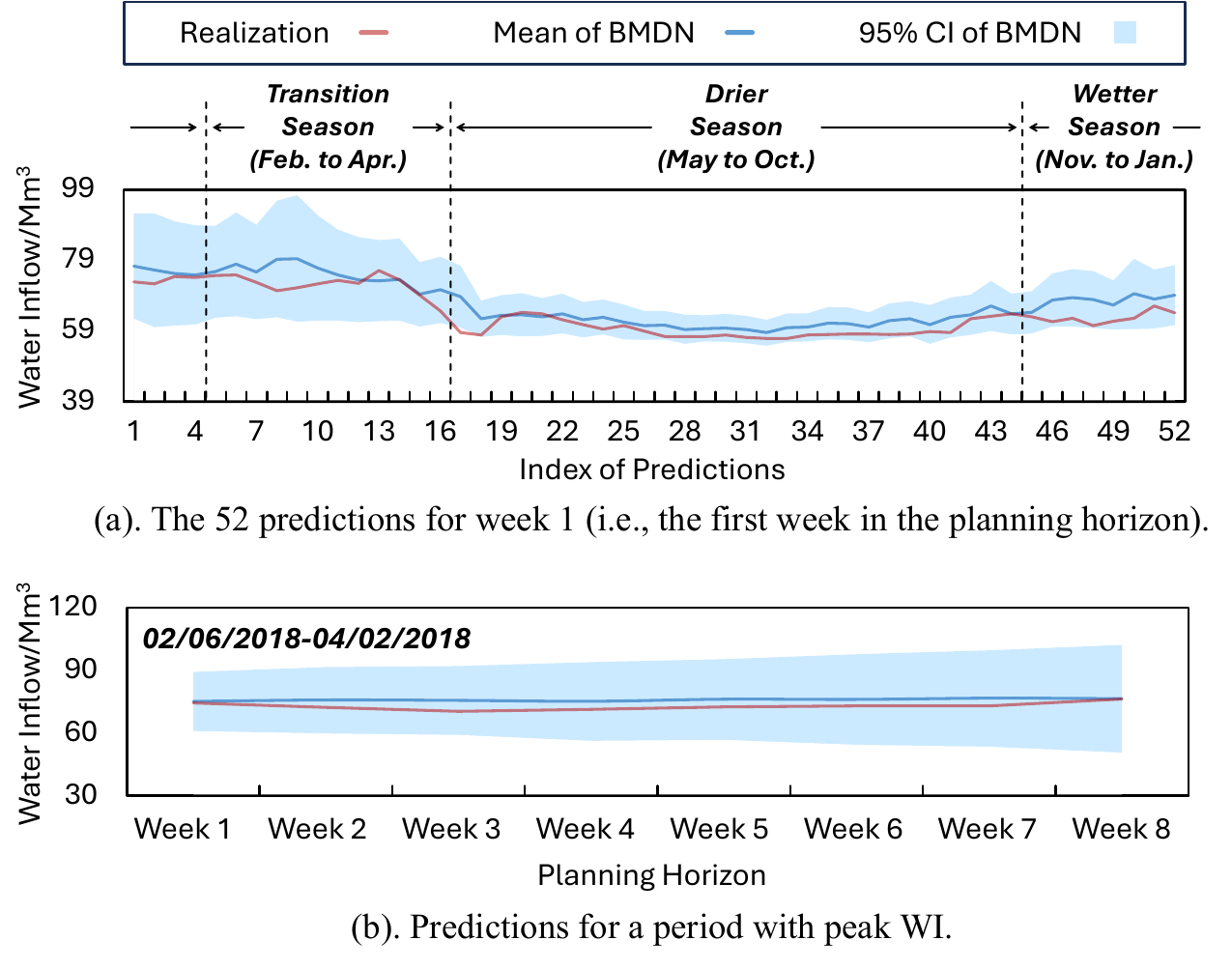}
  \vspace{-7mm}
\caption{Comparison of predictions. The prediction rolls forward weekly. Each prediction result covers the next 8 weeks.}\label{Fig07}
  \vspace{-0mm}
\end{figure}

Fig.~\ref{Fig07} sketches the prediction results of BMDN compared to actual WI realizations, with 95\% confidence intervals (CI) to reflect aleatoric uncertainty. Fig.~\ref{Fig07}(a) compares week 1 WI forecasts from 52 $T+L$ rolling prediction runs along the planning horizon, showing that \textit{i)} BMDN's CIs can cover 51 out of 52 WI realizations; and \textit{ii)} CIs in drier seasons are narrower than wetter seasons with more volatile hydrological conditions. Fig.~\ref{Fig07}(b) shows WI forecasts from one $T+L$ WI prediction run, where nearer weeks generally have a narrower CI, forming a trumpet-shaped blue area. These observations imply that BMDN is more confident in delivering more accurate WI predictions for nearer weeks and in drier seasons.

\vspace{-0mm}
\subsection{Carryover Storage and LMWV-Based Future Value}\label{PGEInsight}
\vspace{-0mm}
\subsubsection{Analysis of Carryover Storage Results}
\begin{figure}[tb]
	\centering
  \vspace{-0mm}
		\includegraphics[width=\columnwidth]{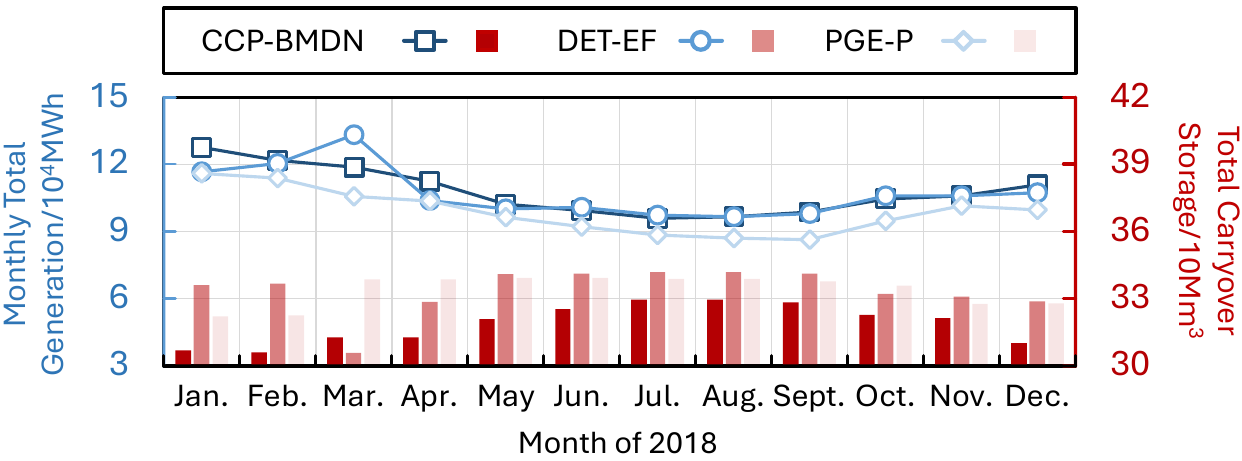}
  \vspace{-7mm}
	\caption{Actual hydropower generation (lines) and carryover storage (bars).}\label{Fig08}
 \vspace{-0mm}
\end{figure}

Fig.~\ref{Fig08} compares the monthly hydropower generation and carryover storage ($V^{\text{cs}}_{\text{RB}} + V^{\text{cs}}_{\text{PT}}$) over 2018 with $L=\text{4}$. According to the PGE dataset, the storage of the RB and PT at the end of 12/31/2017 was respectively {\color{CBlue}319.53 Mm\textsuperscript{3} and 2.85 Mm\textsuperscript{3}}. Regarding the monthly generation, the blue lines show that CCP-BMDN is comparable to DET-EF and outperforms the practical PGE-P.
Indeed, the total hydropower generation in 2018 for CCP-BMDN, DET-EF, and PGE-P is 1,311.81 GWh, 1,303.05 GWh, and 1,201.22 GWh\textemdash CCP-BMDN is 0.67\% higher than DET-EF and 9.21\% higher than PGE-P. 

Moreover, the red bars illustrate that the carryover storage from CCP-BMDN is slightly lower than those of DET-EF and PGE-P. Nevertheless, our experimental results indicate that the carryover storage plans of CCP-BMDN do not trigger any violations of hydropower unit operations. Additionally, it is noteworthy that the average daily generation of DET-EF, which uses error-free predictions, is 0.024 GWh lower than that of CCP-BMDN. This difference arises because BMDN tends to slightly overestimate WI realizations, as shown in Fig.~\ref{Fig07}(a), resulting in a lower future value for the same carryover storage in CCP-BMDN compared to DET-EF. Consequently, CCP-BMDN discharges more water than DET-EF.

Summing the total generation in 2018 and the future value at the end of December 2018, CCP-BMDN, DET-EF, and PGE-P respectively achieve 1,333.8 GWh, 1,334.2 GWh, and 1,226.1 GWh. That is, CCP-BMDN and DET-EF outperform PGE-P by 8.28\% and 8.82\%, respectively.

Fig.~\ref{Fig08} also shows that CCP-BMDN and PGE-P exhibit similar seasonal trends in generation and carryover storage: generation gradually decreases while carryover storage increases from January to September, then generation rises while carryover storage declines for the rest of the year. This similarity implies that CCP-BMDN is as seasonally adaptable as the experience-based PGE-P. These results indicate that CCP-BMDN tends to suitably lower carryover storage to boost generation; despite this, the physical operation constraints can still be guaranteed due to the consideration of uncertainties.

\begin{figure}[tb]
	\centering
   \vspace{-0mm}
		\includegraphics[width=\columnwidth]{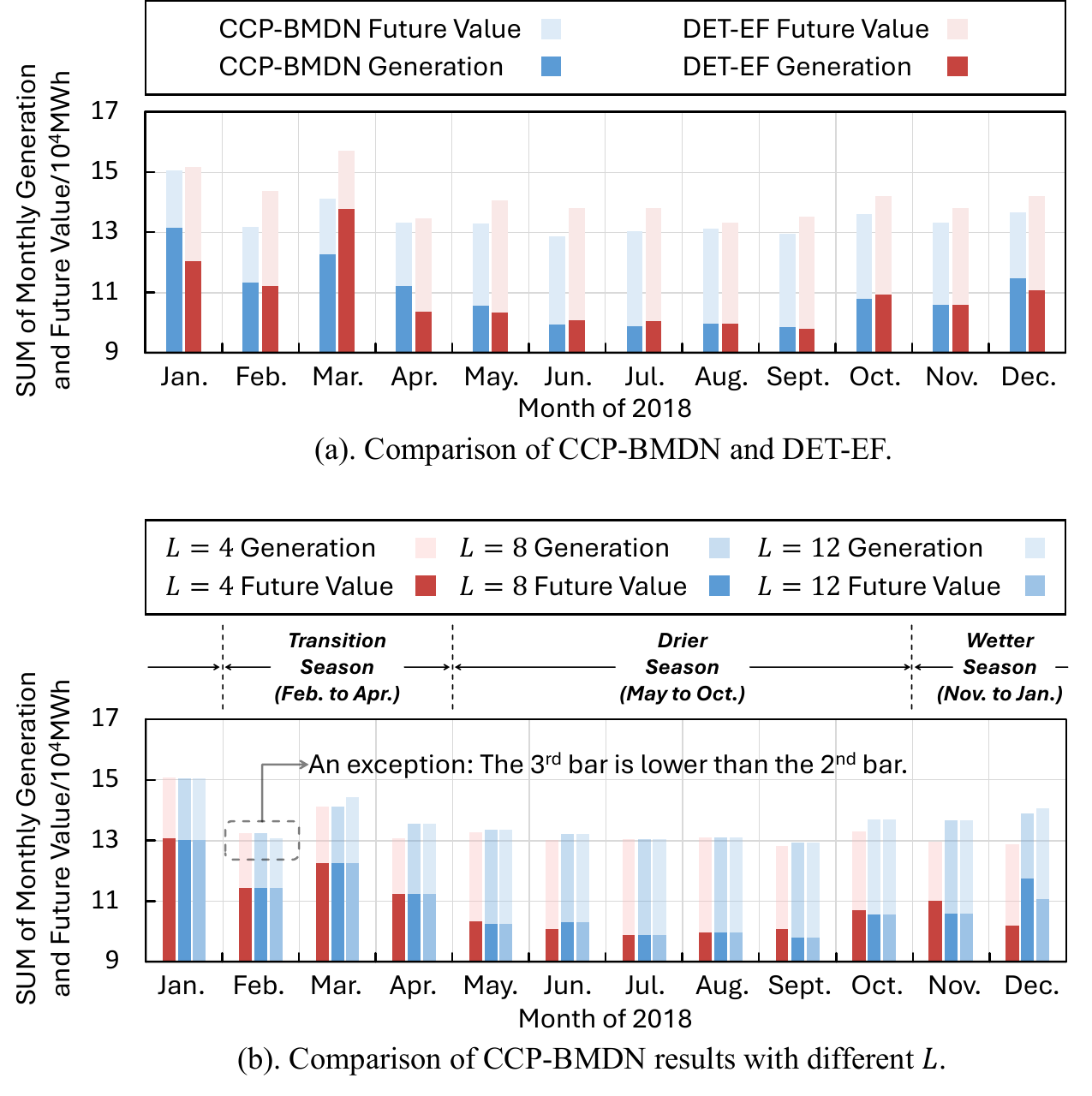}
  \vspace{-8mm}
		\caption{Sum of the actual hydropower generation and evaluated future value.}\label{Fig09}
  \vspace{-3mm}
\end{figure}

Fig.~\ref{Fig09}(a) compares the future value and generation of CCP-BMDN and DET-EF. The sum of the actual monthly generation and evaluated end-of-month future value of CCP-BMDN is generally smaller than that of DET-EF. This gap is expected because CCP-BMDN considers uncertainties to bear with prediction errors. One practical way to narrow this gap is to tune the look-ahead length $L$.

{\color{CBlue}
To this end, Fig.~\ref{Fig09}(b) sketches the results for different values of $L$, showing that except for February, increasing $L$ raises the summed actual monthly generation and the evaluated end-of-month future value. Moreover, our numerical results indicate that \textit{i)} regarding the annual generation, $L=\text{4}/\text{8}/\text{12}$ leads to 1,303.2/1,310.8/1,304.1 GWh, and \textit{ii)} regarding the summed annual generation and end-of-December future value, $L=\text{4}/\text{8}/\text{12}$ leads to 1330.0/1332.1/1333.9 GWh. These findings suggest that properly extending the look-ahead length could improve overall hydropower generation.

The exception in February occurs because the setting of \(L = \text{12}\) extends the future period into May (a drier-season month), whereas February belongs to the transition season. Consequently, LMWVs derived under \(L = \text{12}\) (\(\boldsymbol{\pi} = [\text{490.44}, \text{147.132}]^{\top}\)) exceed those under \(L = \text{8}\) (\(\boldsymbol{\pi} = [\text{441.396}, \text{147.132}]^{\top}\)) and \(L = 4\) (\(\boldsymbol{\pi} = [\text{392.352}, \text{98.088}]^{\top}\)) which remain fully within the transition season. Finally, the setting of \(L = \text{12}\) results in relatively low immediate hydropower generation in February. These sensitivity tests imply that, although a longer planning horizon could potentially enhance long-term water usage, this benefit may not universally apply to all months. In other words, each month has its own optimal \(L\) to ensure the planning horizon properly covers the representative months of the upcoming water season.
}

\subsubsection{Further Analysis of the LMWV-Based Future Value Quantification}
\begin{figure}[tb]
	\centering
		\includegraphics[width=\columnwidth]{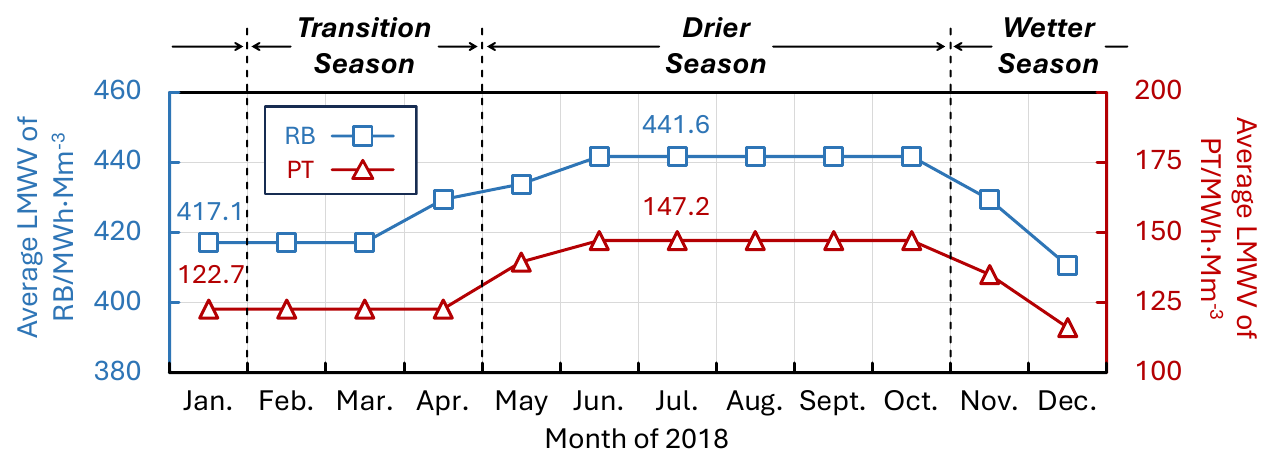}
  \vspace{-7mm}
		\caption{Average locational marginal water values of RB and PT reservoirs.}\label{Fig10}
  \vspace{-0mm}
\end{figure}

Recall that CCP-BMDN offers seasonal adaptability. In fact, this adaptability stems from the derived LMWVs, i.e., the monthly average LMWVs ($\frac{1}{R}\sum_{r=1}^{R}\pi_{r,n}$) in wetter/drier seasons are lower/higher, as shown in Fig.~\ref{Fig10}.

For example, LMWV of RB/PT is {\color{CBlue}417.078/122.670 MWh$\cdot$Mm\textsuperscript{-3}} in January and increases to {\color{CBlue}441.612/147.204 MWh$\cdot$Mm\textsuperscript{-3}} in July. That is, the generation potential of an incremental unit of water in July is higher than in January. This observation aligns with PGE's experience\textemdash for drier/wetter future seasons, LMWVs should be higher/lower because natural WIs are scarce/abundant. Thus, when using LMWVs to weigh up the immediate benefit and future value, higher/lower LMWVs will induce more/less carryover storage, further helping CHPs adapt to upcoming drier/wetter conditions.

\begin{figure}[tb]
	\centering
		\includegraphics[width=\columnwidth]{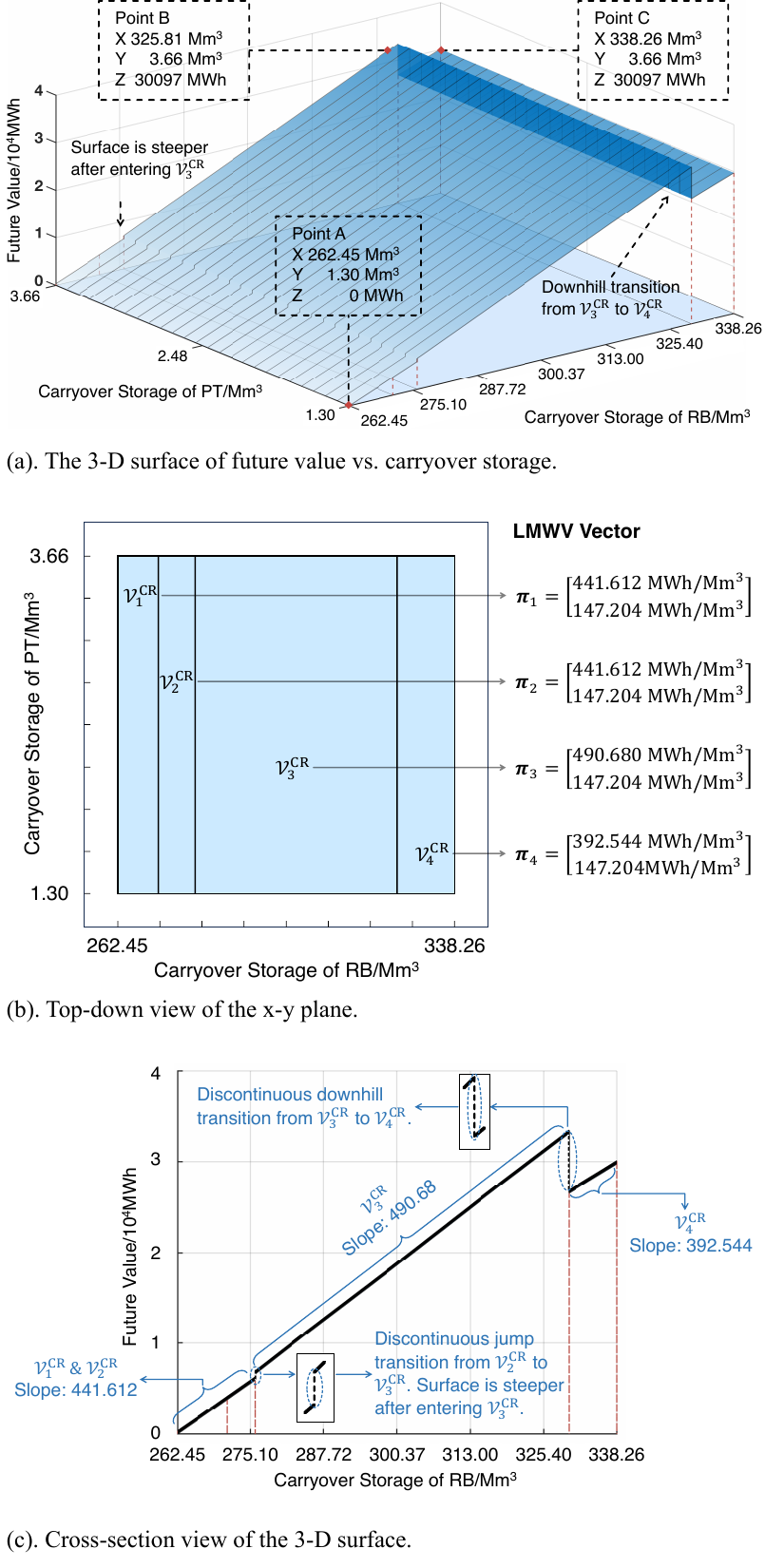}
  \vspace{-7.5mm}
		\caption{Visualization of the evaluated future value of August 2018.}\label{Fig11}
  \vspace{-0mm}
\end{figure}
Another advantage of the presented LMWV-based quantification method is the visualizability of the rules \eqref{Ifthen}, which can assist operators in understanding carryover storage plans. Based on the results of August 2018, Fig.~\ref{Fig11}(a) sketches the relationship between the future value and the carryover storage as a 3-D surface, and the corresponding partition results and the cross-section view are shown in Fig.~\ref{Fig11}(b) and Fig.~\ref{Fig11}(c), respectively. The following six points are noteworthy:
\begin{itemize}
{\color{CBlue}
     \item[{\textit{a).}}] The future value surface presented in Fig.~\ref{Fig11}(a) is complete. This is because the “if-then” rule \eqref{Ifthen} fully covers all possible values of carryover storage variables $[V_{\text{RB}}^{\text{cs}}, V_{\text{PT}}^{\text{cs}}]^{\top}$ within their feasible regions. In contrast, SDP-based methods (e.g., \cite{SDP_Helseth}) must discretize $[V_{\text{RB}}^{\text{cs}}, V_{\text{PT}}^{\text{cs}}]^{\top}$, which inevitably results in certain information loss in the obtained future value surface;     }

     \item[{\textit{b).}}]  Figs.~\ref{Fig11}(a) and \ref{Fig11}(b) show that the feasible region of carryover storage is partitioned into 4 CRs. Each CR is associated with a LMWV vector $\boldsymbol{\pi}_{r}=[\pi_{r,\text{RB}}, \pi_{r,\text{PT}}]^{\top}$;
     
     \item[{\textit{c).}}] Fig.~\ref{Fig11}(a) shows that the future value of Point \textit{A} within $\mathcal{V}_{\text{1}}^{\text{CR}}$ is 0 MWh because its carryover storage is at the lower bound of the storage range;     

     \item[{\textit{d).}}] Fig.~\ref{Fig11}(b) shows that the boundaries of CRs are vertical, indicating that the partition is mainly driven by $V^{\text{cs}}_{\text{RB}}$. Moreover, it is noteworthy that $\pi_{r,\text{PT}}$ are lower than $\pi_{r,\text{RB}}$, as also shown in Fig.~\ref{Fig10}. This is because RB, as a larger reservoir on the upstream, has a higher hydropower generation efficiency; in addition, the upstream water can be re-used in the downstream PT for hydropower generation. Thus, the carryover storage of RB presents a dominating effect on the partition results;

     \item[{\textit{e).}}] Figs.~\ref{Fig11}(a) and \ref{Fig11}(c) show that the surface extends upward with $\boldsymbol{\pi}_{r}$ as the gradient, and is steeper after transitioning from $\mathcal{V}_{\text{2}}^{\text{CR}}$ to $\mathcal{V}_{\text{3}}^{\text{CR}}$. This transition is due to a change of binary variable solutions (e.g., some units are turned on after entering $\mathcal{V}_{\text{3}}^{\text{CR}}$). The slope becomes steeper because $\pi_{\text{3,RB}} \geq \pi_{\text{2,RB}}$;
     
    \item[{\textit{f).}}] Figs.~\ref{Fig11}(a) and \ref{Fig11}(c) show a downhill transition from $\mathcal{V}_{\text{3}}^{\text{CR}}$ to $\mathcal{V}_{\text{4}}^{\text{CR}}$ with $\pi_{r,\text{RB}}$ dropping by {\color{CBlue}98.136 MWh$\cdot$Mm\textsuperscript{-3}}. This implies that simply increasing carryover storage could reduce LMWVs, as the most efficient operating condition occurs in $\mathcal{V}_{\text{3}}^{\text{CR}}$. The CCP-BMDN model captures this insight well: it identifies Point \textit{B} as the optimal carryover storage plan for August, positioned near the highest point of $\mathcal{V}_{\text{3}}^{\text{CR}}$, so as to prepare the CHP for the dry September;

{\color{CBlue}
\item[{\textit{g).}}] 
Fig.~\ref{Fig11}(c) shows that the future value surface is discontinuous: an abrupt jump occurs when transitioning from $\mathcal{V}_{\text{2}}^{\text{CR}}$ to $\mathcal{V}_{\text{3}}^{\text{CR}}$, and a sharp drop appears when moving from $\mathcal{V}_{\text{3}}^{\text{CR}}$ to $\mathcal{V}_{\text{4}}^{\text{CR}}$. These discontinuities arise because model \eqref{FModel} includes binary variables (e.g., $I_{ni}$), and each CR corresponds to a distinct solution of these binary variables (i.e., $\boldsymbol{y}^{\star}_r$ in Fig.~\ref{Fig04}). As the carryover storage $\boldsymbol{V}^{\text{cs},\theta}$ transits from one CR to another, the changes in the binary solutions lead to discontinuous jumps in the objective function of model \eqref{FModel}. Since the future value evaluated via LMWVs reflects the changes in the objective value \eqref{FModel:1} with respect to the carryover storage parameters, it naturally inherits these discontinuities. Unlike SDDP-based methods (e.g., \cite{Arild1, Arild_aggregation, Arild6, Arild7, Arild_TPS, binaryexpansion}), which rely on convex hyperplanes and cannot fully capture these non-convex characteristics, the proposed quantification framework precisely represents these discontinuities. As a result, our framework provides CHP operators with a complete geometric understanding of how the future value changes across the entire feasible operation region.}
\end{itemize}

Our experiments also indicate that the performance of the partition-then-extract algorithm varies by season: compared to drier seasons, partitioning for wetter seasons generally takes less time and leads to fewer CRs\textemdash 5.58 seconds vs. 15.01 seconds, and 3 CRs vs. 5 CRs. This is due to the abundant natural WIs during wetter seasons, which causes near-full storage throughout the season. As a result, most hydropower units stay online, and only a few of them will switch their ON-OFF statuses within the parameter space, leading to fewer CRs (see Definition~\ref{CR}) and fewer iterations for the algorithm.

{\color{CBlue}
\subsection{Applying the Partition-Then-Extract Algorithm on an 8-Reservoir CHP}

\begin{table}[tb]
{\color{CBlue}	\caption{Results on the 8-Reservoir CHP}\label{Tab03}
	\centering
	\footnotesize
\begin{tabular}{ccc}
\toprule  
             \multirow{2}{*}{WI/Mm\textsuperscript{3}}   & Partition & \multirow{2}{*}{Avg. LMWVs of Reservoirs/10\textsuperscript{2}MWh$\cdot$Mm\textsuperscript{-3}}  \\
             & Time/s &   \\
\midrule
            303.37& 185.82 &  [9.36, 6.50, 5.52, 4.58, 3.64, 2.70, 1.80, 0.90]\\
\midrule
            692.49& 14.82  &  [8.42, 6.17, 5.32, 4.42, 3.52, 2.66, 1.76, 0.90]\\
\bottomrule
\end{tabular}
}
\end{table}

\begin{figure}[tb]
{\color{CBlue}	\centering
 \vspace{-0mm}
		\includegraphics[width=\columnwidth]{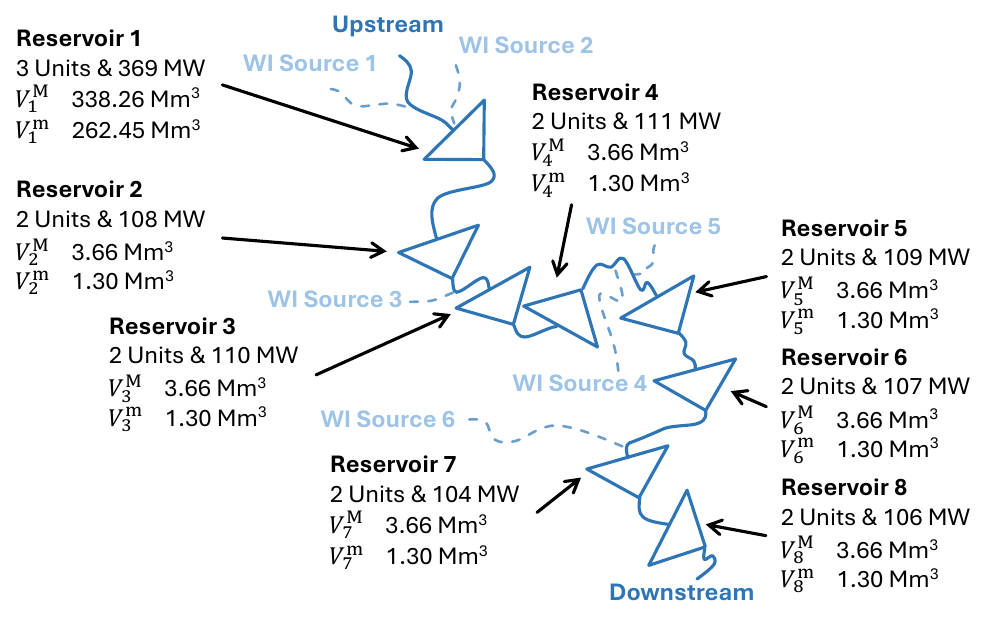}
  \vspace{-4mm}
		\caption{Illustration of the 8-reservoir CHP.}\label{Fig12}
   \vspace{-0mm}}
\end{figure}

A larger system modified from the PT-RB CHP, as shown in Fig.~\ref{Fig12}, is further used to assess the scalability of the presented framework. Two cases are tested: a drier season case with an expected total WI of 303.37 Mm\textsuperscript{3} over the 4-week future period (i.e., $\sum_{n=1}^{8}\sum_{l=T+1}^{T+4} \hat{W}^\text{fp}_{nl}=\text{303.37 Mm\textsuperscript{3}}$), and a wetter season case with 692.49 Mm\textsuperscript{3}. Overall, the conclusions drawn from the PGE case are found to be extendable to this larger CHP, as highlighted in Table~\ref{Tab03}:
\begin{itemize}
     \item
     \textit{Computational Time:}
The partition-then-extract algorithm is significantly more time-consuming in the drier season compared to the wetter season (185.82 seconds vs. 14.82 seconds). Additionally, compared to the PGE case, the partition times for wetter seasons are similar (14.82 seconds vs. 5.58 seconds), but are notably different for drier seasons (185.82 seconds vs. 15.01 seconds). This suggests that hydrological conditions could strongly influence the scalability of the partition-then-extract algorithm for large-scale CHPs;
     
     \item
     \textit{Seasonal LMWV Trend:}
LMWVs in the wetter season are slightly lower than those in the drier season, indicating that the hydrological adaptivity of the algorithm remains valid;

    \item
    \textit{Geographic LMWV Trend:}
LMWVs decrease gradually from upstream to downstream, implying that water stored in upstream reservoirs can contribute more significantly to the future value.
\end{itemize}}

\vspace{-0mm}
\section{Conclusions}\label{Sec5}
\vspace{-0mm}
To support medium-term CHP planning decision-making, this paper presents a framework to quantify the hydropower generation potential of carryover storage and tests its effectiveness on a CHP operated by PGE. The resulting analytical quantification rules can be directly integrated into medium-term CHP planning models as tractable linear constraints, assisting CHP operators in balancing immediate benefit and future value. More importantly, these rules offer straightforward physical interpretations to aid operators in understanding the quantification results. Numerical results based on practical data indicate that deploying the quantification rules in a CCP-based medium-term CHP planning model can improve hydropower generation for PGE by 9.21\%.

{\color{CBlue}
As the presented future-period model only considers a single deterministic scenario, a possible avenue for future work is to extend it to a stochastic counterpart that accounts for future water inflow uncertainties while retaining visualization advantages. Another promising direction is to further enhance the scalability of the framework for larger-scale CHPs under more complex setups, such as additional technical details (e.g., head dependencies, responses to electricity prices, and operational characteristics with greater chronological detail and/or finer time resolution) and operational intricacies (e.g., following instructions of independent system operators).
}

\begin{figure}[tb]
{\color{CBlue}
	\centering
 \vspace{-0mm}
		\includegraphics[width=\columnwidth]{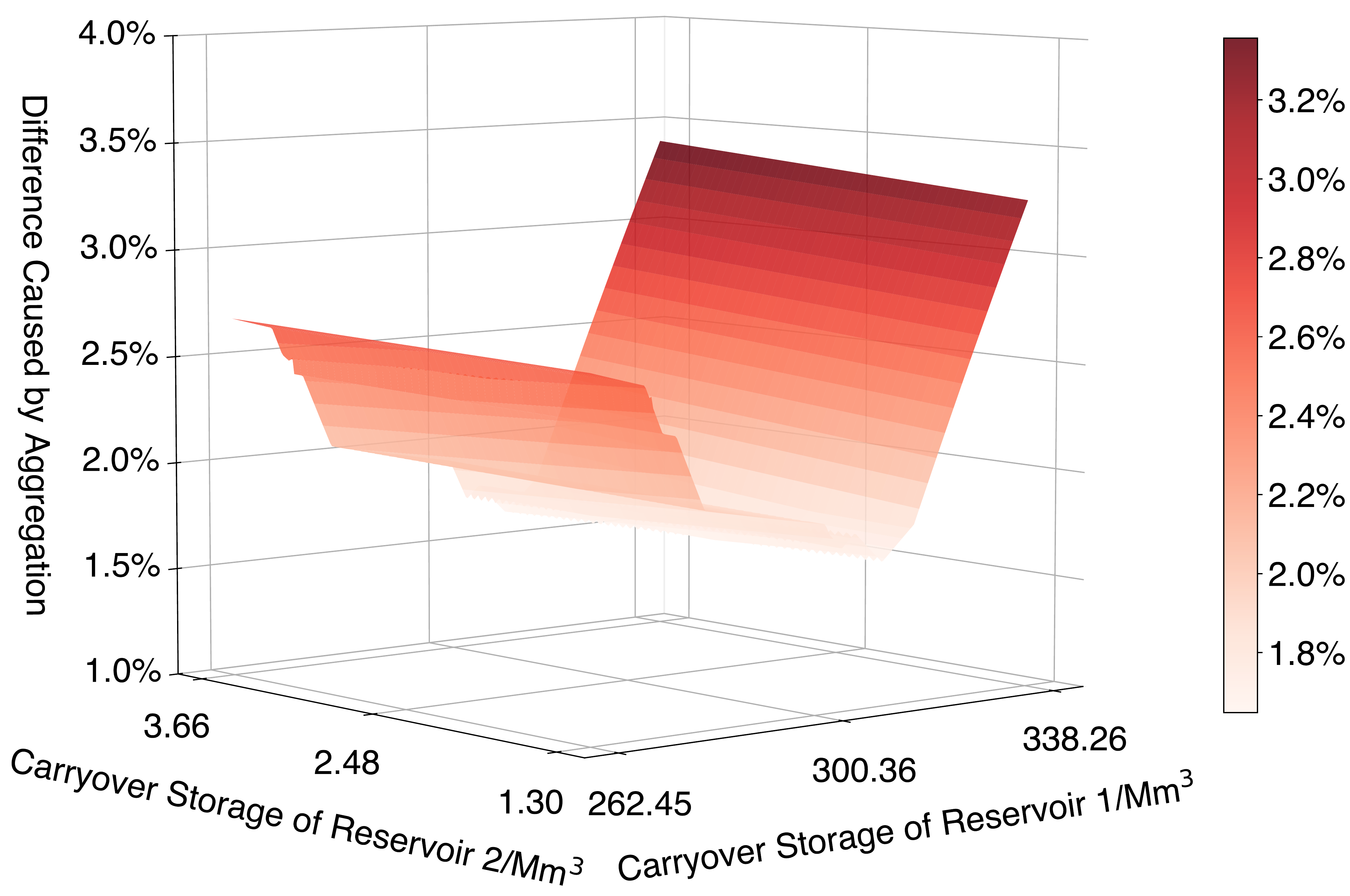}
  \vspace{-7mm}
	\caption{Objective value differences caused by the aggregation.}\label{Fig13}
	\vspace{-0mm}}
\end{figure}

\appendix
{\color{CBlue}
\subsection{The Impact of Aggregation on Objective Values of \eqref{FModel}} \label{gap}

Model \eqref{EX} describes the full-model counterpart of \eqref{FModel} without aggregation, which preserves the time-step index \(l\). This subsection compares the objective values of the aggregated model \eqref{FModel} and the full model \eqref{EX}.

\begin{subequations}\label{EX}
\begin{flalign}
&\textstyle{\max\limits_{\Psi}
\sum\nolimits_{n \in \mathcal{N}} \sum\nolimits_{i \in \mathcal{I}_{n}} \sum\nolimits_{l \in \mathcal{L}}\lambda P_{nil}}  
-  \sum\nolimits_{n \in \mathcal{N}} \sum\nolimits_{l \in \mathcal{L}} C^{\text{ws}}_{n} S_{nl}
                                 \mspace{-150mu}& \notag     \\
&\text{where } \Psi=\{
       \boldsymbol{D},
       \boldsymbol{I},
       \boldsymbol{P},
       \boldsymbol{S},
       \boldsymbol{V},
       \boldsymbol{W}^{\Delta}\}                                                   \mspace{-150mu}&  \label{EX:1} \\
&\text{s.t. }\textstyle{V^{\text{m}}_{n} \leq V_{nl} \leq V^{\text{M}}_{n},}          \mspace{-150mu}&\forall n, \forall l;    \label{EX:2}\\
&\mspace{27mu}\textstyle{V_{n,l+1}=V_{n}^{\text{cs},\theta}+\sum\nolimits_{\tau=T+1}^{l}(\hat{W}_{n\tau} + W^{\Delta}_{n\tau}),}                                                                     \mspace{-150mu}&\forall n, \forall l;    \label{EX:3}\\
&\mspace{27mu}\textstyle{W^{\Delta}_{nl}=\sum\nolimits_{m \in \bar{\mathcal{N}}_{n}}(\sum\nolimits_{i \in \mathcal{I}_{m}} \alpha D_{m,i,l-\delta} + S_{m,l-\delta}) }                                                       \mspace{-150mu}&   \notag\\
&\mspace{27mu}\textstyle{\quad -\sum\nolimits_{i \in \mathcal{I}_{n}} \alpha D_{nil} - S_{nl},\,S_{nl} \geq 0,}              \mspace{-150mu}&\forall n, \forall l;   \label{EX:4}\\
&\mspace{27mu}\textstyle{P_{nil} = \mathcal{P}^{\text{RtP}}(D_{nil},I_{nil}),I_{nil} \in \{0, 1\},}                                                                \mspace{-150mu}&\forall n, \forall i, \forall l;\label{EX:5}\\
&\mspace{27mu}\textstyle{P^{\text{m}}_{ni} I_{nil} \leq P_{nil} \leq P^{\text{M}}_{ni} I_{nil},}                       \mspace{-150mu}&\forall n, \forall i, \forall l;\label{EX:6}\\
&\mspace{27mu}\textstyle{D^{\text{m}}_{ni} I_{nil} \leq D_{nil} \leq D^{\text{M}}_{ni} I_{nil},}
                                                                 \mspace{-150mu}&\forall n, \forall i, \forall l;\label{EX:7}
\end{flalign}
\end{subequations}

The comparison is performed on the PGE CHP case shown in Fig.~\ref{Fig05}. In practice, PGE restricts the CHP forebay levels within a narrow range, inducing the storage ranges of [262.45, 338.26] Mm\textsuperscript{3} for $V_{1}^{\text{cs},\theta}$ and [1.30, 3.66] Mm\textsuperscript{3} for $V_{2}^{\text{cs},\theta}$. We set \(\hat{W}^\text{fp}_{n} = \sum_{\tau=T+1}^{L} \hat{W}_{n\tau}\) and \(L = T+4\). A total of 12,500 combinations of \(\bigl(V_{1}^{\text{cs},\theta}, V_{2}^{\text{cs},\theta}\bigr)\) is sampled, and the difference in objective values of \eqref{FModel} and \eqref{EX} is calculated as \eqref{difference}. Fig.~\ref{Fig13} shows the differences in objective values resulting from the aggregation. All differences range from 1.0\% to 3.5\%, with an average of 2.1\%. These tests suggest that although aggregation slightly increases the objective value, the impact remains acceptable, considering the computational benefits of model \eqref{FModel}.
\begin{equation}\label{difference}
\text{Difference} = \frac{\text{Obj \eqref{FModel:1} $-$ Obj \eqref{EX:1}}}{\text{Obj \eqref{EX:1}}}  \times 100\%
\end{equation}

}

\subsection{The CCP-BMDN Medium-Term CHP Planning Model} \label{DetailedModel1}
The medium-term planning model of CCP-BMDN is formulated as in \eqref{MainModel}, which is an enhanced counterpart of \eqref{General} that also considers WI prediction uncertainties of the current period. The superscript $\diamond$ is used to distinguish current period variables from future period ones. 

The objective function \eqref{MainModel:1} is to maximize hydropower generation during the current and future periods by optimizing carryover storage. The joint chance constraint (JCC) \eqref{MainModel:JCC} ensures that storage limits are satisfied with a probability of at least $\text{1}-\epsilon_{t}$ under uncertain WIs, while \eqref{MainModel:3} calculates the water volume difference between inflows and outflows. The delay time $\delta$ of WI is set to 0 according to the PGE operation manual. Each reservoir $n$ is associated with a WI prediction $\hat{\boldsymbol{W}}_{n}^{\text{cp}}$ provided by BMDN, which is described by the GMM \eqref{GMM}. The $t^{\text{th}}$ element of $\hat{\boldsymbol{W}}_{n}^{\text{cp}}$, denoted as $\hat{W}^{\text{cp}}_{nt}$, corresponds to the WI prediction for week $t$ \cite{Arild5}.

Constraint \eqref{MainModel:4} calculates the storage evolution under the expected WI prediction $\hat{W}_{nt}^{\text{cp}, \mu}$, where $\hat{W}_{nt}^{\text{cp}, \mu}$ is the mean vectors of the GMM \eqref{GMM}; \eqref{MainModel:5} limits the storage volume; \eqref{MainModel:6} defines the carryover storage as the storage level at the beginning of week $T+1$ under the expected WI prediction; \eqref{MainModel:7}-\eqref{MainModel:9} have the same meaning as \eqref{FModel:7}-\eqref{FModel:9}, with additional consideration of week indices; and \eqref{MainModel:10} is the linear constraints expressing the ``if-then'' rules \eqref{Ifthen}.
\begin{subequations}\label{MainModel}
\begin{flalign}
&\textstyle{\max\limits_{\Xi, \boldsymbol{V}^{\text{cs}}}
\sum\nolimits_{n \in \mathcal{N}} \sum\nolimits_{i \in \mathcal{I}_{n}} \sum\nolimits_{t \in \mathcal{T}}\lambda P^{\diamond}_{nit}  + F(\boldsymbol{V}^{\text{cs}})}                                   \mspace{-150mu}& \notag     \\
&\text{where } \Xi=\{
       D_{nit}^{\diamond},
       I_{nit}^{\diamond},
       P_{nit}^{\diamond},
       S_{nt}^{\diamond},
       V_{n}^{\diamond},
       W^{\Delta}_{nt},
       Z_{r}\}                                                   \mspace{-150mu}&  \label{MainModel:1} \\
&\text{s.t. } \mathbb{P}\textstyle{\left\{ \begin{array}{l}
V_{n,1}^{\diamond}+\sum\nolimits_{\tau=1}^{t} (\hat{W}_{n\tau}^{\text{cp}} + W^{\Delta}_{n\tau}) \leq V^{\text{M}}_{n}, \forall n;\\
V_{n,1}^{\diamond}+\sum\nolimits_{\tau=1}^{t} (\hat{W}_{n\tau}^{\text{cp}} + W^{\Delta}_{n\tau}) \geq V^{\text{m}}_{n}, \forall n;
\end{array} \right\} \geq 1 - \epsilon_{t},}                     \mspace{-150mu}&      \notag\\
&                                                                \mspace{-150mu}&\forall t;      \label{MainModel:JCC}\\
&\mspace{25mu}\textstyle{W^{\Delta}_{nt}=\sum\nolimits_{m \in \bar{\mathcal{N}}_{n}}(\sum\nolimits_{i \in \mathcal{I}_{m}} \alpha D^{\diamond}_{m,i,t-\delta} + S^{\diamond}_{m,t-\delta}) }                                                       \mspace{-150mu}&   \notag\\
&\mspace{25mu}\textstyle{\quad -\sum\nolimits_{i \in \mathcal{I}_{n}} \alpha D^{\diamond}_{nit} - S^{\diamond}_{nt},\,S_{nt}^{\diamond} \geq 0,}              \mspace{-150mu}&\forall n, \forall t;   \label{MainModel:3}\\
&\mspace{25mu}\textstyle{V^{\diamond}_{n,t+1}=V^{\diamond}_{n,1}+\sum\nolimits_{\tau=1}^{t}(\hat{W}_{n\tau}^{\text{cp}, \mu} + W^{\Delta}_{n\tau}),}                                                                     \mspace{-150mu}&\forall n, \forall t;    \label{MainModel:4}\\
&\mspace{25mu}\textstyle{V^{\text{m}}_{n} \leq V^{\diamond}_{nt} \leq V^{\text{M}}_{n},}
                                                                 \mspace{-150mu}&\forall n, \forall t;    \label{MainModel:5}\\
&\mspace{25mu}\textstyle{V_{n}^{\text{cs}}=V_{n,T+1}^{\diamond},}\mspace{-150mu}&\forall n;               \label{MainModel:6}\\
&\mspace{25mu}\textstyle{P^{\diamond}_{nit} = \mathcal{P}^{\text{RtP}}(D^{\diamond}_{nit},I^{\diamond}_{nit}),I_{nit}^{\diamond} \in \{0, 1\},}
                                                                 \mspace{-150mu}&\forall n, \forall i, \forall t; \label{MainModel:7}\\
&\mspace{25mu}\textstyle{P^{\text{m}}_{ni} I_{nit} \leq P_{nit}^{\diamond} \leq P^{\text{M}}_{ni} I_{nit},}
                                                                 \mspace{-150mu}&\forall n, \forall i, \forall t; \label{MainModel:8}\\
&\mspace{25mu}\textstyle{D^{\text{m}}_{ni} I_{nit} \leq D_{nit}^{\diamond} \leq D^{\text{M}}_{ni} I_{nit},}
                                                                 \mspace{-150mu}&\forall n, \forall i, \forall t; \label{MainModel:9}\\
&\mspace{25mu}\textstyle{\text{Calculation rules of future value: \eqref{Easytouse}};}   
                                                                 \mspace{-150mu}&                \label{MainModel:10}
\end{flalign}
\end{subequations}

A JCC \eqref{MainModel:JCC} for week $t$ is then converted into deterministic constraints by the following three steps:

\textbf{Step 1.} For week $t$, apply Boole's inequality \cite{Boole} to split its JCC \eqref{MainModel:JCC} into 2$N$ individual chance constraints (ICCs) \eqref{ICC}. Each ICC is associated with a probability $\epsilon_{nt}^{\text{M}}$ or $\epsilon_{nt}^{\text{m}}$.
\begin{flalign}\label{ICC}
&\textstyle{\mathbb{P} \{V^{\text{M}}_{n}\mspace{-2mu}-\mspace{-2mu}V^{\diamond}_{n,1}\mspace{-2mu}-\mspace{-2mu}\sum\nolimits_{\tau=1}^{t}W^{\Delta}_{n\tau} \geq \sum\nolimits_{\tau=1}^{t}}\hat{W}_{n\tau}^{\text{cp}} \}\mspace{-2mu} \geq\mspace{-2mu} 1 \mspace{-2mu}-\mspace{-2mu} \epsilon_{nt}^{\text{M}}, \forall n;\mspace{-30mu}& \notag\\
&\textstyle{\mathbb{P} \{V^{\text{m}}_{n}\mspace{-2mu}-\mspace{-2mu}V^{\diamond}_{n,1}\mspace{-2mu}-\mspace{-2mu}\sum\nolimits_{\tau=1}^{t}W^{\Delta}_{n\tau} \leq \sum\nolimits_{\tau=1}^{t}}\hat{W}_{n\tau}^{\text{cp}} \}\mspace{-2mu} \geq\mspace{-2mu} 1 \mspace{-2mu}-\mspace{-2mu} \epsilon_{nt}^{\text{m}}, \,\forall n;\mspace{-30mu}&
\end{flalign}

The ICCs \eqref{ICC} are safe approximations to the JCC \eqref{MainModel:JCC} if $\sum_{n=1}^{N}(\epsilon_{nt}^{\text{M}}+\epsilon_{nt}^{\text{m}}) \leq \epsilon_{t}$ holds \cite{Boole}. In this paper, $\epsilon^{\text{M}}_{nt}$ and $\epsilon_{nt}^{\text{m}}$ are set as $\epsilon_{t}/\text{2}\textit{N}$; $\epsilon_{t=\text{1}}$, $\epsilon_{t=\text{2}}$, $\epsilon_{t=\text{3}}$, and $\epsilon_{t=\text{4}}$ are set as 0.010, 0.015, 0.020, and 0.025, respectively, reflecting higher probability guarantees for more recent weeks. With these, the 2$N$ ICCs \eqref{ICC} can be equivalently reformulated as quantile constraints \eqref{Quantile}, in which $\mathcal{Q}^{\epsilon_{t}}$ is the $\epsilon_{t}$ quantile function.
\begin{flalign}\label{Quantile}
&V^{\text{M}}_{n}-V^{\diamond}_{n,1} - \textstyle{\sum\nolimits_{\tau=1}^{t}}W^{\Delta}_{n\tau} \geq
\textstyle{\mathcal{Q}^{1-(\epsilon_{t}/\text{2}\textit{N})}(\sum\nolimits_{\tau=1}^{t}}\hat{W}_{n\tau}^{\text{cp}}), \forall n; &\notag\\
&V^{\text{m}}_{n}-V^{\diamond}_{n,1} - \textstyle{\sum\nolimits_{\tau=1}^{t}}W^{\Delta}_{n\tau} \leq
\textstyle{\mathcal{Q}^{\epsilon_{t}/\text{2}\textit{N}}(\sum\nolimits_{\tau=1}^{t}}\hat{W}_{n\tau}^{\text{cp}}), \mspace{31mu} \forall n; \mspace{-30mu}    &
\end{flalign}

\textbf{Step 2.} Rewrite $\sum_{\tau=1}^{t}\hat{W}^{\text{cp}}_{n\tau}$ in \eqref{Quantile} as $\boldsymbol{s}_{nt}^{\top}\hat{\boldsymbol{W}}^{\text{cp}}_{n}$, where $\boldsymbol{s}_{nt}$ is a constant vector with 0s and 1s at appropriate places. The affine invariance of GMM is then leveraged to replace $\boldsymbol{s}_{nt}^{\top}\hat{\boldsymbol{W}}^{\text{cp}}_{n}$ with a 1-dimensional variable, denoted as $\hat{W}^{\prime}_{nt}$ \cite{Wu_UC}. The problem now is converted to calculating the quantiles $\mathcal{Q}^{1-(\epsilon_{t}/\text{2}\textit{N})}(\hat{W}_{nt}^{\prime})$ and $\mathcal{Q}^{\epsilon_{t}/\text{2}\textit{N}}(\hat{W}_{nt}^{\prime})$, which are indeed the roots of the univariate nonlinear equations \eqref{root}.
\begin{equation}\label{root}
\textstyle{\text{CDF}_{\hat{W}_{nt}^{\prime}}(\rho) = 1 - ({\epsilon_{t}}/{\text{2}\textit{N}}),\,\text{CDF}_{\hat{W}_{nt}^{\prime}}(\rho) = {\epsilon_{t}}/{\text{2}\textit{N}}, \, \forall n;}
\end{equation}

\textbf{Step 3.}
Apply Newton method to solve \eqref{root}. Denote the results as $\rho^{\epsilon_{t}/\text{2}\textit{N}}$ and $\rho^{1-(\epsilon_{t}/\text{2}\textit{N})}$. Rewrite the quantile constraints \eqref{Quantile} as the deterministic linear constraints \eqref{Quantile_New}.
\begin{flalign}\label{Quantile_New}
&V^{\text{M}}_{n}-V_{n,1}-\textstyle{\sum\nolimits_{\tau=1}^{t}}W^{\Delta}_{n\tau}\geq \rho^{1-(\epsilon_{t}/\text{2}\textit{N})}, \mspace{5mu}\forall n;\notag\\
&V^{\text{m}}_{n}-V_{n,1}-\textstyle{\sum\nolimits_{\tau=1}^{t}}W^{\Delta}_{n\tau}\leq \rho^{\epsilon_{t}/\text{2}\textit{N}},     \mspace{36mu} \forall n;
\end{flalign}

After applying the above three steps for weeks $t=1,...,T$, the original $T$ JCCs \eqref{MainModel:JCC} can be replaced with 2$NT$ tractable reformulations \eqref{Quantile_New}, finally converting model \eqref{MainModel} into a deterministic MILP that can be solved by MILP solvers.

\subsection{The DET-EF Medium-Term CHP Planning Model} \label{DetailedModel2}
\vspace{-0mm}
The medium-term planning model of DET-EF is a deterministic variant of \eqref{MainModel}, which replaces $\hat{W}^{\text{cp},\mu}_{n\tau}$ in \eqref{MainModel:4} with error-free predictions and thus naturally degrades the JCC \eqref{MainModel:JCC} into a deterministic constraint.

{\color{CBlue}
\subsection{The Short-Term Scheduling Model to Assess the Determined Target Carryover Storage Level} \label{STModelforTest}
The short-term scheduling model \eqref{ST}, as a deterministic counterpart of \eqref{MainModel:1}–\eqref{MainModel:9}, is used to evaluate the effectiveness of the determined target carryover storage level. In this model, \(\tilde{{W}}_{nt}\) denotes the WI realizations, and the right-hand-side parameter \({V}_{n}^{\text{cs}}\) in constraint \eqref{ST:target} is set as the determined target carryover storage level. The objective value \eqref{ST:1} describes the actual hydropower generation (i.e., \textit{immediate benefit}) over $T$ weeks.
\begin{subequations}\label{ST}
\begin{flalign}
&\textstyle{\max\limits_{\Phi}
\sum\nolimits_{n \in \mathcal{N}} \sum\nolimits_{i \in \mathcal{I}_{n}} \sum\nolimits_{t \in \mathcal{T}}\lambda P^{\diamond}_{nit}}                                   \mspace{-150mu}& \notag     \\
&\text{where } \Phi=\{
       D_{nit}^{\diamond},
       I_{nit}^{\diamond},
       P_{nit}^{\diamond},
       S_{nt}^{\diamond},
       V_{n}^{\diamond},
       W^{\Delta}_{nt}\}                                                   \mspace{-150mu}&  \label{ST:1} \\
&\text{s.t. } \textstyle{V^{\text{m}}_{n} \leq V^{\diamond}_{nt} \leq V^{\text{M}}_{n},}          \mspace{-150mu}&\forall n, \forall t;    \label{ST:2}\\
&\mspace{27mu}\textstyle{W^{\Delta}_{nt}=\sum\nolimits_{m \in \bar{\mathcal{N}}_{n}}(\sum\nolimits_{i \in \mathcal{I}_{m}} \alpha D^{\diamond}_{m,i,t-\delta} + S^{\diamond}_{m,t-\delta}) }                                                       \mspace{-150mu}&   \notag\\
&\mspace{27mu}\textstyle{\quad -\sum\nolimits_{i \in \mathcal{I}_{n}} \alpha D^{\diamond}_{nit} - S^{\diamond}_{nt},\,S_{nt}^{\diamond} \geq 0,}              \mspace{-150mu}&\forall n, \forall t;   \label{ST:3}\\
&\mspace{27mu}\textstyle{V^{\diamond}_{n,t+1}=V^{\diamond}_{n,1}+\sum\nolimits_{\tau=1}^{t}(\tilde{W}_{n\tau} + W^{\Delta}_{n\tau}),}                                                                     \mspace{-150mu}&\forall n, \forall t;    \label{ST:4}\\
&\mspace{27mu}\textstyle{V_{n,T+1}^{\diamond}=V_{n}^{\text{cs}},}\mspace{-150mu}&\forall n;               \label{ST:target}\\
&\mspace{27mu}\textstyle{P^{\diamond}_{nit} = \mathcal{P}^{\text{RtP}}(D^{\diamond}_{nit},I^{\diamond}_{nit}),I_{nit}^{\diamond} \in \{0, 1\},}
                                                                 \mspace{-150mu}&\forall n, \forall i, \forall t;\label{ST:6}\\
&\mspace{27mu}\textstyle{P^{\text{m}}_{ni} I_{nit} \leq P_{nit}^{\diamond} \leq P^{\text{M}}_{ni} I_{nit},}
                                                                \mspace{-150mu}&\forall n, \forall i, \forall t;\label{ST:9}\\
&\mspace{27mu}\textstyle{D^{\text{m}}_{ni} I_{nit} \leq D_{nit}^{\diamond} \leq D^{\text{M}}_{ni} I_{nit},}
                                                                 \mspace{-150mu}&\forall n, \forall i, \forall t;\label{ST:8}
\end{flalign}
\end{subequations}
}
\bibliographystyle{IEEEtran}
\bibliography{Hydro_Refs}

\begin{IEEEbiographynophoto}{Xianbang Chen} (Student Member, IEEE) received the B.S. and M.S. degrees in electrical engineering from Sichuan University, Chengdu, China, in 2017 and 2020, respectively, and the Ph.D. degree in electrical engineering from Stevens Institute of Technology (SIT), Hoboken, NJ, USA, in 2024. He is currently a postdoctoral fellow at the Electric Power and Energy Systems Laboratory at SIT. His research interests focus on enhancing the operational efficiency of modern power systems through hybrid optimization and learning methodologies.
\end{IEEEbiographynophoto}

\begin{IEEEbiographynophoto}{Yikui Liu} (Member, IEEE) received a B.S. degree in electrical engineering and automation from the Nanjing Institute of Technology, China, in 2012, the M.S. degree in power system and automation from Sichuan University, China, in 2015, and the Ph.D. degree in electrical and computer engineering from the Stevens Institute of Technology, Hoboken, NJ, USA, in 2020. He worked at Siemens, USA, as an energy market engineer from 2020 to 2021. He worked as a Postdoctoral Researcher with the Stevens Institute of Technology, Hoboken, NJ, USA, from 2021 to 2023. He is currently an associate researcher at Sichuan University, Chengdu, China. His research interests include the power market and OPF in distribution systems.
\end{IEEEbiographynophoto}

\begin{IEEEbiographynophoto}{Zhiming Zhong} received a Ph.D. degree in Industrial Engineering and Operations Research at the University of Arizona in 2024. Prior to that, he received the M.S. and B.S. degrees in Technical Economics and Management Science at North China Electric Power University, respectively, in 2020 and 2017. His research interests include optimization and decision-making under uncertainty, large-scale optimization, and their applications in power system scheduling and the electricity market.
\end{IEEEbiographynophoto}

\begin{IEEEbiographynophoto}{Neng Fan} received a Ph.D. degree in industrial and systems engineering from the University of Florida, Gainesville, FL, USA, in 2011. He is currently an Associate Professor with the University of Arizona, Tucson, AZ, USA. His research interests include mathematical optimization, data analytics, and their applications to power systems reliability, and renewable energy integration.
\end{IEEEbiographynophoto}

\begin{IEEEbiographynophoto}{Zhechong Zhao} received a B.S. degree in electrical engineering and automation from the Shanghai University of Electric Power, Shanghai, China, in 2008, the M.S. degree in electrical engineering from the Illinois Institute of Technology, Chicago, IL, USA, in 2011, and the Ph.D. degree in electrical engineering from Clarkson University, Potsdam, NY, USA, in 2014. He worked at Portland General Electric, where he was responsible for developing and implementing trading strategies to optimize a diverse 4GW portfolio of assets in the Pacific Northwest.
\end{IEEEbiographynophoto}

\begin{IEEEbiographynophoto}{Lei Wu} (Fellow, IEEE) received the B.S. degree in electrical engineering and the M.S. degree in systems engineering from Xi’an Jiaotong University, Xi’an, China, in 2001 and 2004, respectively, and the Ph.D. degree in electrical engineering from Illinois Institute of Technology (IIT), Chicago, IL, USA, in 2008. From 2008 to 2010, he was a Senior Research Associate with the Robert W. Galvin Center for Electricity Innovation, IIT. He was a summer Visiting Faculty at NYISO in 2012. He was a Professor with the Electrical and Computer Engineering Department, Clarkson University, Potsdam, NY, USA, till 2018. Currently, he is Anson Wood Burchard Chair Professor with the Department of Electrical and Computer Engineering, Stevens Institute of Technology, Hoboken, NJ, USA. His research interests include power systems operation and planning, energy economics, and community resilience microgrid.\end{IEEEbiographynophoto}

\end{document}